%% file: ServiceProtocols.tex
\documentclass[final]{elsarticle}

\newcommand{\myTitle}{A Formalization of Social Requirements  for Human Interactions with Service Protocols}

\input{envPackages}
\input{envCommands}

\input{envTheorems}

\journal{Information Sciences}

\begin{document}
\input{frontMatter}

\input{section_Introduction}
\input{section_Rationale}
\input{section_RelatedWork}
\input{section_Background}

%\input{tex/section_Requirements}
\input{section_FormalModel}

\input{section_Applications}
\input{section_Discussion}
\input{section_Conclusions}

%\section*{References}
\bibliography{bibliography}
\bibliographystyle{model1-num-names}

\end{document}

%% file: envPackages.tex
\usepackage[greek, english]{babel}
\usepackage{paralist}%
\usepackage{amsmath}%
\usepackage{amsfonts}%
\usepackage{fourier}
\usepackage{MnSymbol}
\usepackage{graphicx}
\usepackage[T1]{fontenc}

\usepackage{url}

\usepackage{courier}

\usepackage{xspace}

%% file: envCommands.tex
\newcommand{\backbullet}{\ensuremath{\,\,^{\bullet}\!\!}}
\newcommand{\topbullet}{\ensuremath{\,\,\,^{^{\bullet{}}}\!\!\!\!\!\!\!\!}}

\newcommand{\topalpha}{\ensuremath{\,\,\,^{^{\alpha{}}}\!\!\!\!\!\!\!}}
\newcommand{\aLambda}{\ensuremath{\topalpha\Lambda}}

\newcommand{\ie}{i.e., }
\newcommand{\eg}{e.g., }

\newcommand{\klas}[1]{#1$^\alpha$}

\newcommand{\AT}{\texttt{Archi\-bald} \texttt{Tex}\xspace}
\newcommand{\ATtext}{Archi\-bald Tex\xspace}
\newcommand{\DevHouse}{Dev\-House}
\newcommand{\DHit}{\emph{\DevHouse}\xspace}
\renewcommand{\DH}{\DevHouse\xspace}
\newcommand{\DHtt}{\texttt{\DevHouse}\xspace}
\newcommand{\DevBank}{\texttt{De\-vel\-op\-er\-Bank}\xspace}
\newcommand{\Collab}{\texttt{Col\-lab\-o\-ra\-tion}\xspace}
\newcommand{\MoniBankText}{MoniBank\xspace}
\newcommand{\MoniBank}{\texttt{\MoniBankText}\xspace}

\newcommand{\ExpArchClass}{\klas{\texttt{Ex\-pe\-ri\-enced} \texttt{Ar\-chi\-tect}}\xspace}
\newcommand{\ExpDevClass}{\klas{\texttt{Ex\-pe\-ri\-enced} \texttt{De\-vel\-op\-er}}\xspace}
\newcommand{\BankClass}{\klas{\texttt{Bank}}\xspace}
\newcommand{\DevBankClass}{\klas{\texttt{De\-vel\-op\-er\-Bank}}\xspace}
\newcommand{\CollabClass}{\klas{\texttt{Col\-lab\-o\-ra\-tion}}\xspace}
\newcommand{\SitePrepClass}{\klas{\texttt{Site} \texttt{Pre\-pa\-ra\-tion}}\xspace}
\newcommand{\RecommendClass}{\klas{\texttt{Recommend}}\xspace}

%% file: envTheorems.tex
\newcounter{defNumber}

\newenvironment{definition}[1]
{
\medskip
\refstepcounter{defNumber}\noindent\textbf{Definition~\arabic{defNumber} (#1)}\hspace{1ex}}
{
\smallskip
}

%% file: frontMatter.tex
\begin{frontmatter} 
\title{\myTitle{}}

\author[UEP]{Willy Picard}
\ead{picard@kti.ue.poznan.pl}
\address[UEP]{Pozna\'{n} University of Economics,
Department of Information Technology, \\
al. Niepodleg\l{}o\'{s}ci 10, 61-875 Pozna\'{n}, 
Poland}

\begin{abstract}
Collaboration models and tools aim at improving the efficiency and effectiveness of human interactions.  Although social relations among collaborators have been identified as having a strong influence on collaboration, they are still insufficiently taken into account in current collaboration models and tools. In this paper, the concept of service protocols is proposed as a model for human interactions supporting social requirements, \ie sets of constraints on the relations among interacting humans. Service protocols have been proposed as an answer to the need for models for human interactions in which not only the potential sequences of activities are specified -- as in process models -- but also the constraints on the relations among collaborators. 
Service protocols  are based on two main ideas: 
first, service protocols are rooted in the service-oriented architecture (SOA): each service protocol contains a \emph{service-oriented summary} which
provides a representation of the activities of an associated process model in SOA terms.
 Second, a class-based graph---referred to as a \emph{service network schema}---restricts the set of potential service elements that may participate in the service protocol by defining constraints on nodes and constraints on arcs, \ie social requirements.
Another major contribution to the modelling of human interactions is a unified approach organized around the concept of service, understood in a broad sense with services being not only Web services, but also provided by humans.
\end{abstract}


\end{frontmatter}

%% file: section_Introduction.tex
\section{Introduction}
An important goal of many information systems is to support human interactions. The spectrum of information systems focusing on human interactions is very broad, from critical systems such as air traffic control systems, to most modern video games, such as World of Warcraft. Among enterprise information systems, both enterprise resource planning (ERP) and customer relationship management (CRM) management systems aim at supporting human interactions.

The development of service-oriented architecture (SOA) and business process management (BPM) has lead to a new approach to the design and implementation of information systems. In this approach, information systems are designed and implemented around the concept of decoupled and orchestrated processing entities, referred to as \emph{services}.

Currently, SOA is mainly applied at the infrastructure level, with most development efforts related to the implementation of SOA with Web services and WS-* standards. Business Process Modelling in the context of SOA leads to the orchestration of Web services for which two standards are prevalent: WS-BPEL~\cite{standard:BPEL20} for Web services orchestration, and BPMN~\cite{standard:OMG:BPMN12} for more business-oriented and higher-level models.

The importance of human interactions in the context of SOA and BPM has been lately recognized. Emerging standards, such as WS-BPEL Extension for People (usually shortened to BPEL4People)~\cite{standard:BPEL4People} and WS-HumanTask~\cite{standard:WS-HumanTask}, aim at providing better support for activities performed by humans in the BPEL framework. However, these two standards do not recentre BPEL around human-to-human interactions but rather propose a formal definition of human activities and potential inclusion of these tasks within a BPEL process. As a consequence, information systems developed in the SOA paradigm are usually supporting human interactions in a limited and insufficient manner.

%Two main reasons for the poor computer support for human interactions in the context of SOA and BPM are \emph{adaptation capabilities} of humans and \emph{social elements} involved in the interactions among humans. Adaptation capabilities refer to the capability of humans to react to changes in their environment. Adaptation capabilities of humans are not only far more advanced than adaptation capabilities of software entities, but also not taken into account in existing models for collaboration processes.

The main reason for the poor computer support for human interactions in the context of SOA and BPM is the difficulty to model \emph{social elements} involved in the interactions among humans. 
An attempt to encompass social elements involved in the interactions among humans is the introduction of the concept of a role. A role  usually defines the right to perform a limited set of activities within a given process. Depending on one's social position or competences, various roles may be attributed. Another approach, popular in social networking websites such as Facebook~\cite{www:facebook} or LinkedIn~\cite{www:linkedin}, consists in linking individuals and organizations with a set of predefined types of relations, such as ``Friend'' or ``Connections''.

Neither roles nor predefined types of relations reflect the complexity of social elements that play an important role in the interactions among humans. Some examples of such social elements are positions within a given hierarchical structure of an organization, trust relations, past cooperation, and recommendations among humans.

%A proper computer support for human interactions requires an appropriate handling of social elements. In the area of BPM, social elements are mainly \emph{social requirements}. In this paper, a social requirement is defined as a set of constraints on the relations among interacting actors, \ie requirements about social relations among actors.

Although many information systems support human interactions, they do not encompass social elements. As a consequence, there is a mismatch between the support for human interactions provided by the information systems and the social elements, part of the social norm that is ruling the interactions among the humans using the information system. Among various manifestations of this mismatch, social networking websites are often violating rules defined in the social norm, leading to legal and ethical issues. An example is the possibility on Facebook to publish a photography about a person without the consent of this person: such a situation is usually condemned by social norms as a privacy violation, except in the case when the allowed viewers of the photography are friends or close family. Facebook support for human interactions could be improved by taking into consideration the relations between the photographed person and the viewers, \ie social aspects, allowing only friends and close family to access the photography.

Therefore, there is a need for novel models of human interactions, especially models encompassing social elements, to develop information systems supporting human interactions in a more efficient and effective manner. The concepts of service and process model that underlies SOA and BPM should be extended to encompass social aspects and their influence on the human interactions.

In this paper, a formalization of social requirements for human interactions is proposed as a novel model of human interactions supporting social aspects is proposed. Social requirements are defined as sets of constraints on relations among interacting actors. In the presented model, referred to as \emph{service protocols}, a process model defining actors and their potential sequences of activities is extended by a set of constraints concerning both the actors and their relations. As an example, while a model for the process of house building defines potential activities for the architect, for the building company, and for a gardener, a service protocol may ensure that the architect and the building company have a long cooperation history and that the building company and the gardener have the same suppliers.

The rest of this paper is organized as follows. First, a rationale for service protocols is presented in Section~\ref{sec:rationale}. In Section~\ref{sec:relatedWork}, related work is presented. Next, some basic notions related to classes, graphs, processes, and services are presented. Then, a formal model of service protocols is presented in Section~\ref{sec:formalModel}. Potential applications of service protocols are proposed in Section~\ref{sec:applications}. A discussion of service protocols is presented in Section~\ref{sec:discussion}. Finally, Section~\ref{sec:conclusions} concludes this paper.

%% file: section_Rationale.tex
\section{Rationale for Service Protocols}
\label{sec:rationale}

In this section, the rationale for service protocol is presented. First, a running illustrative example grounded in the construction sector is introduced. Next, based on the running example, a set of key issues is identified. Then, the goals of service protocols are presented. Finally, a list of requirements for service protocols is proposed.

\subsection{An Illustrative Example}
\label{subsec:example}

To introduce the main key issues that service protocols aim at addressing, a running example grounded in the construction sector is presented in this section. For the sake of readability of this paper, the presented example is much simpler than real cases.

Let assume that a \emph{real-estate developer company}, named \DHit, is planning its next investment. \DH currently owns a field in which an old abandoned oil refinery is standing in ruins. \DH plans to build a supermarket in this field. \DH is an experienced real-estate developer with 24 investments.

First, \DH should obtain a loan from a \emph{bank} to finance the investment. To ease the whole procedure, \DH wants to focus on banks with which it has already collaborated. Therefore, \DH plans to negotiate a loan with banks at which it currently has an account, and from which it has got former loans. However, \DH would like to avoid a next loan at a bank from which it already has a current loan. Finally, banks proposing interest rates higher than 5.5\% for a 3-year loan should be rejected. A potential candidate for a bank financing \DH's project is \emph{\MoniBankText}. \MoniBankText is a commercial bank with about 235.500 clients, offering a 3\% interest rate for 3-year loans. \DH has an account and no current loan at \MoniBankText.

Second, \DH needs an \emph{architect} to develop the construction plans. A major requirement concerning the architect concerns past and current collaboration. As the supermarket is a strategically important investment for \DH, the architect should have at least five former projects performed for \DH, but less than two projects currently performed. A potential candidate for the architect is a Canadian architect, named \emph{\ATtext}, responsible for 17 investments. \ATtext collaborates with \DH on 3  current projects, following a collaboration on 15 past projects. 

Third, \DH needs the field to be cleaned. The developer does not want to be in charge of the supervision of the clearance process. The architect should be responsible for identifying a \emph{site preparation company}, which will supervise the preparation tasks, \ie demolition and rubble removal. The two companies that will perform the preparation tasks, \ie a \emph{demolition company} and a \emph{debris hauling company}, should be known and trusted by the architect. Additionally they should be able to efficiently collaborate.
%\begin{figure}[htp]
%	\centering
%		\includegraphics[width=330pt]{figs/TypedSchema}
%	\caption{TO BE DELETED}
%\end{figure}

The remaining tasks required to build a supermarket, \eg foundation construction, masonry, carpentry, are not taken into account in this paper.

\subsection{Key Issues}
Although a classical case in the construction sector is described in the formerly presented illustrative example, a set of issues related to this case are still to be addressed.

\subsubsection{Multiple Service Consumers}
In a traditional orchestration approach, various service pro\-vi\-ders are provisioning the service interfaces that, assembled together, form a process. As an example, a BPEL process model usually relies on various service providers, but only one service consumer exists: the BPEL engine that is executing the process instance.

In the formerly presented illustrative example, as well as in many cases of collaboration among companies, a process consists not only of various service pro\-vi\-ders, but also various service consumers.
In the example, a first service consumer is \DH, the real-estate developer, that is seeking for financing from banks. Next, a second service consumer is the architect that needs a site preparation company to supervise the cleaning of the field. Finally, the site preparation company itself is a service consumer when it consumes the demolition and rubble removal services provided by the demolition company and the debris hauling company.

Not only multiple service providers, but also multiple service consumers should therefore be encompassed in models for collaboration. As a consequence, collaboration could be modelled as a set of collaborators---the service consumers---per\-for\-ming activities provided by actors---the service providers.
%---that may eventually be outside the set of collaborators. The bank is an example of such a service provider that is outside the set of collaborators in the formerly presented illustrative example.
 
\subsubsection{Constraints on Actors}
In a traditional BPM approach, process models define constraints on the potential sequences of tasks and the concept of role is used to limit the execution of a given task to individuals with the appropriate rights. In short, a process model defines \emph{who may do what, and when}. However, it should be noted that the ``who'' part, \ie the role definition, is usually limited to a label associated with a set of tasks that \emph{may} be performed. A process role defines a set of \emph{rights}. 

In the formerly presented illustrative example, a process model may define that each individual playing the \texttt{Real-estate Developer} role may perform the \texttt{ne\-go\-tiate a loan} task. A real-estate developer may negotiate a loan if financing is needed. However, if the investment may be fully financed by the real-estate developer, a loan (and the associated negotiations to obtain it) may be avoided.

Another type of constraints may be identified: some tasks may be performed only by actors with appropriate characteristics.  Although process roles define rights, constraints on actors define \emph{obligations}. The obligations associated with a given actor may be considered as a definition of the ``who'' part formerly mentioned.

In the formerly presented illustrative example, a constraint on an actor being a \texttt{Real-estate Developer} may state that her \texttt{number of investments} has to be greater than ten. Any real-estate developer with a lesser number of investments should not be allowed to participate in the formerly presented process. Similarly, banks proposing interest rates higher than 5.5\% for a 3-year loan should be rejected.

Although constraints on actors are often concerning their competences to perform a given task\footnote{As an example, architects usually have to complete an appropriate examination to be licensed, such as the Architect Registration Examination (ARE) for Canadian architects.}, constraints on actors may concern a multitude of aspects of actors, e.g, their physical traits\footnote{In a soccer team, the goalkeeper should rather be tall.} or their place of living\footnote{In election-related processes, the place of voting is usually related to the place of living.}.

\subsubsection{Relational Constraints}
A second type of constraints may be identified: relational constraints. Some tasks may be performed only by actors with appropriate relations with other actors. Similarly to constraints on actors, relational constraints define \emph{obligations}. However, while the constraints on actors focus on the essential characteristics of an actor, \ie the characteristics of the actor in isolation, the obligations defined by relational constraints focus on the environment of the actor and her social place in this environment.

In the formerly presented illustrative example, the choice of the architect is limited by relational constraints: the architect should have at least five former projects performed for \DH, but less than two projects currently performed. Similarly, the choice of the bank is limited by relational constraints: \DH is interested only in banks at which it has a bank account, former loans should have been obtained from the bank by \DH, and currently \DH should not have any loan from the bank.

\subsection{Goals of Service Protocols}

First, service protocols aim at providing a model for \emph{service-based collaboration encompassing the multiplicity of service consumers}. Traditional process models, \eg BPEL, are usually based on the assumption that, although the responsibility for activity execution is spread among actors, only one actor---usually the process engine---is responsible for activity invocations. In SOA terms, traditional process models assume a single service consumer and various service providers. In many collaboration processes, \eg the formerly presented illustrative example, various service consumers are collaborating. Service protocols tackle the problem of collaboration processes in which the responsibility for activity invocations is spread among various actors, \ie various service consumers.

The second goal of service protocols is to support the \emph{definition of constraints on actors}, both service providers and service consumers. In traditional process models, actors are usually related to roles. A role provides actors that play it with the right to perform a set of activities in given states of the collaboration process. Constraints on actors are a means to define the obligations that an actor has to fulfil to participate in the collaboration process.

The third goal of service protocols is to support the \emph{definition of relational constraints between actors}.
Although the importance of social aspects in collaboration processes has been largely studied~\cite{book:buchanan:2007}, traditional process models still lack support for relational constraints. Service protocols provide support not only for constraints on actors, but also for the constraints concerning the relations between them. As a consequence, although traditional process models focus on possible sequences of activities associated with a set of roles, service protocols extend traditional process models by providing means to define constraints concerning the group of actors that may execute a given process model.

\subsection{Requirements for Service Protocols}
\label{sec:requirements}

Based on the set of requirements presented in~\cite{picard:prove:2008}, the following requirements for service protocols supporting human interactions in SOA may be articulated:
\begin{enumerate}
	\item \emph{reusability}: a given service protocol should be reusable to rule the interactions within various groups of collaborators; A service protocol aims at modelling a set of collaboration processes, in the same way as a class models a set of objects in object-oriented programming. In other words, a service protocol may be seen as a model whose instances are collaboration processes;
	\item \emph{separation of activities implementation from service protocols}: a service protocol should model potential interactions among collaborators, however the interactions should be decoupled from implementation of the activities performed by collaborators. As a consequence, activities of a given service protocol may be implemented in various ways, using various technologies, or various locations/hosts;
	\item \emph{strong mathematical foundations}: service protocols model complex cases of potential interactions among humans. Therefore, strong mathematical foundations are required as a mean to check properties such as structural validity, reachability, liveness and boundedness of the proposed models;
		\item \emph{support for social aspects in collaboration}: human interactions are strongly related to social aspects. Social aspects may limit the choice of collaborators, impose some relations among interacting humans. Most process modelling languages and notations do not provide designers of process models with means to explicitly capture social requirements concerning collaborators in the process model. Computer support for human interactions should treat social requirements, \ie constraints on the relations among actors of human interactions, as an integral part of the model of human interactions.
\end{enumerate}

With regard to~\cite{picard:prove:2008}, although the three first requirements have just been adapted to service protocols and the SOA context, the fourth requirement has been added to stress the importance of social requirements expressed as constraints on the relations among actors of human interactions.

%% file: section_RelatedWork.tex
\section{Related Work}
\label{sec:relatedWork}

\subsection{Business Process Modeling in SOA}

In the BPM literature, information required to model and control a process has been classified according to various perspectives. 
In~\cite{article:vda:2003}, five perspectives have been presented:
\begin{itemize}
	\item the \emph{functional} perspective focuses on activities to be performed,
	\item the \emph{process} perspective focuses on the execution conditions for activities,
	\item the \emph{organization} perspective focuses on the organizational structure of the population that may potentially execute activities,
	\item the \emph{information} perspective focuses on data flow among tasks,
	\item the \emph{operation} perspective focuses on elementary operations performed by applications and resources.
\end{itemize}

The introduction of BPEL4People and WS-HumanTask may be considered as an attempt to address the organization perspective. However, the proposed solution is rather an attempt to integrate human interaction in business processes composed by Web services. BPEL4People and WS-HumanTask do not address various key aspects of human interactions: Mendling et al. have identified a set of limitations of BPEL4Peo\-ple concerning the important issue of separation of duty in human interaction~\cite{proceedings:mendling:2007}. Similarly, Russell and van der Aalst have shown the boundaries of the expressiveness of BPEL4People and WS-HumanTask, detailing a set of workflow resource patterns that the two standards do not implement~\cite{report:russell:2007}.

Holmes et al. have remarked that BPEL4People and WS-HumanTask are the result of rapid changes in technologies associated with SOA~\cite{proceedings:holmes:2010}. As a consequence, they propose a model-driven approach to business processes, including basic support for human interactions. The proposed model-driven approach is based on simple assumptions about human interactions, especially concerning roles. However, social requirements are missing in the presented model.

\subsection{Computer Support for Human Interactions}

Computer support for human interactions has been the subject of research in various research communities, from computer support for collaborative work (CSCW) and workflow management systems to BPM and adaptive case management (ACM).

CSCW and workflow management systems focus mainly on the handling of documents in a processing chain to which various persons are participating. CSCW and workflow management systems are adopting a document-routing approach in which social aspects are not addressed.
These systems have been studied and developed mainly in the 1990's, with a major result being the publication in 1995 of the Workflow Reference Model~\cite{standard:wfmc:referenceModel} by the Workflow Management Coalition (WfMC). 

In a workflow approach, it is assumed that a process may be designed and implemented once and for all. Most workflow management system were based on this disputable assumption. Next, business process management techniques have been developed to support changing process models. A direct consequence of this shift from workflows to business processes is the later development of SOA based on Web services and appropriate languages to model business processes, \eg BPEL and BPMN.

Swenson has shown that, although BPM focuses on predicable processes, a large number of processes are by nature unpredictable~\cite{book:swenson:2010}. He coined the term Case Management (ACM) that refers to a new approach to support knowledge work, \ie work which is not repeated, unpredictable, emergent and robust in the face of variable conditions. As an example, the work of a fire rescue crew member is usually not repeated as each fire is different, is unpredictable as the situation on the fireplace may hardly be foreseen, is emergent as the immediate work of the crew member is determined by recently discovered knowledge about the fire situation, and robust in the face of variable conditions as the fire rescue crew member should be reliable and perform efficiently whatever the fire situation found.

In ACM, a fundamental idea is that a process may not be modelled a priori (as in BPM), but the design of a process model should be performed at run-time, while the human interactions are ruled by the process model. A similar idea has been largely studied by  Harrison-Broninski~\cite{book:harrinsonBroninski:2005}. Harrison-Broninski has identified that many human interactions are based on flexible, innovative, collaborative human activity. As a consequence, Harisson-Broninski has proposed to define a new class of systems, Human Interaction Management Systems (HIMS), which would support flexibility, innovation and collaboration in human interactions. Although social relations are mentioned a few times in~\cite{book:harrinsonBroninski:2005} as a potential information that may be relevant for flexibility and innovation, the weakest point of both Swenson's and Harrison-Broninski's works remains the lack of clearly articulated solutions to the very well presented set of issues.

Finally, the concept of Social BPM has been recently proposed and is the subject of various research efforts, with associated events such as the 3\textsuperscript{rd} Workshop on Business Process Management and Social Software~\cite{conf:BPMS2'10} at the 8\textsuperscript{th} International Conference on Business Process Management (BPM2010). However, among various definitions of Social BPM~\cite{sap:socialBPM:2009,slides:richardson:2010}, a consensus seems to appear around the idea that Social BPM is about the use of social media and Web 2.0 approach to the design of process model. Therefore, in the Social BPM approach, wikis and blogs may be used to co-author a process model, allowing collaborators to design in a collaborative manner the model for the process within which they are collaborating. In any case, the concept of social requirements as part of the process model is associated with Social BPM.

\subsection{Social Requirements}

A preliminary remark concerning social requirements is that, although the term is widely used, it is usually not defined. As an example, social requirements are directly mentioned in~\cite{article:ackerman:2000} without the provision of a definition.

A definition of social requirements may be found in~\cite{article:swierzowicz:2009}, where social requirements and social network analysis are associated as follows: ``Social Network Analysis may be used to examine a given network by evaluating some of its properties. Social requirements may be considered as the reverse approach: social requirements may be used to define some properties of a network and their associated expected values, that may then be used to check if an existing network satisfies these social requirements.'' A definition of the relation between social requirements and process model is however still missing, as well as a formal definition of social requirements.

%% file: section_Background.tex
\section{Background}
\label{sec:background}
In this section, the concepts related to object-oriented graphs are defined. Next, the main concepts related to processes and services in the SOA approach are defined. These notions are fundamental for a good understanding of the concept of a service protocol presented in Section~\ref{sec:formalModel}.

\subsection{Object-oriented Graphs}
We define two kinds of object-oriented graphs: object-based graphs and class-based graphs.

\subsubsection{Object-based Graphs}

An \emph{object} is a set of properties $o=\{p\}$. A \emph{property} $p$ is a pair $\langle n,v_n \rangle$, where $n$ is the name of the property and $v_n$ is the value of the property. The value of a property may be a literal or an object.

\begin{figure}[htp]
	\centering
		\includegraphics[scale=0.6]{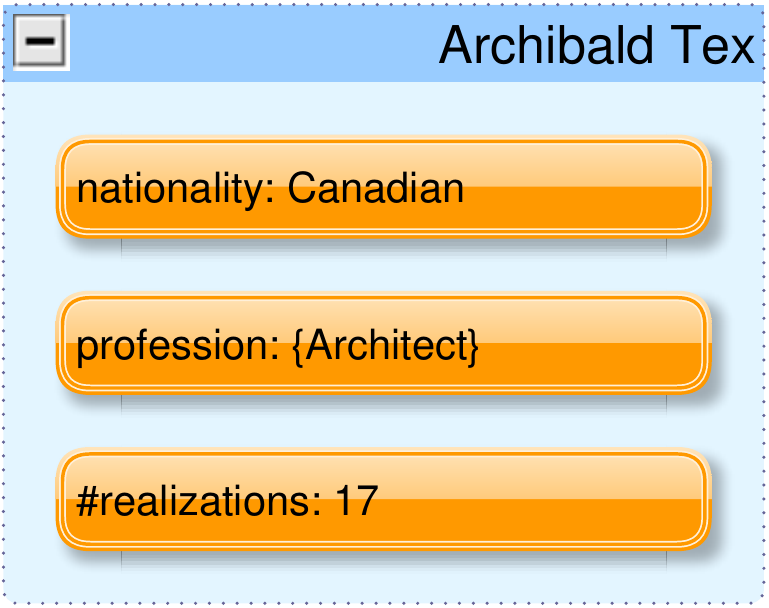}
	\caption{An example of an object}
	\label{fig:ObjectAT}
\end{figure}

As an example, consider the object \AT, illustrated in Figure~\ref{fig:ObjectAT} and defined by the following set of properties:
\begin{itemize}
	\item $\langle\; \texttt{nationality, Canadian}\; \rangle$, 
	\item $\langle\; \texttt{profession, \{Architect\}}\; \rangle$, and 
	\item $\langle\; \texttt{\#realizations}$\footnote{The symbol \# should be understood as ``number of''. Therefore, ``\#realizations'' should be understood as the number of realizations. }$\texttt{, 17}\; \rangle$.
\end{itemize}
In Figure~\ref{fig:ObjectAT}, the external rectangle represents the object \AT, while each inner rounded rectangle represents a property of this object.

A set of objects and their relations may be modelled as an \emph{object-based graph}. 

\begin{definition}{Object-based Graph }
An object-based graph $g= \langle N,\vec{A}\rangle$ is a graph whose nodes $n \in N$ are objects, and arcs connecting nodes $\vec{a} \in \vec{A}$ consist of a source object, a destination object, and an object describing the arc itself.
\end{definition}

To illustrate the concept of a object-based graph, consider a second object defining a real-estate developer \DHtt by the following set of properties:
\begin{itemize}
	\item $\langle\; \texttt{name, DevHouse}\; \rangle$, 
	\item $\langle\; \texttt{profession, \{Real-estate Developer\}}\; \rangle$, and 
	\item $\langle\; \texttt{\#investments, 24}\; \rangle$.
\end{itemize}

\begin{figure}[htp]
	\centering
		\includegraphics[scale=0.6]{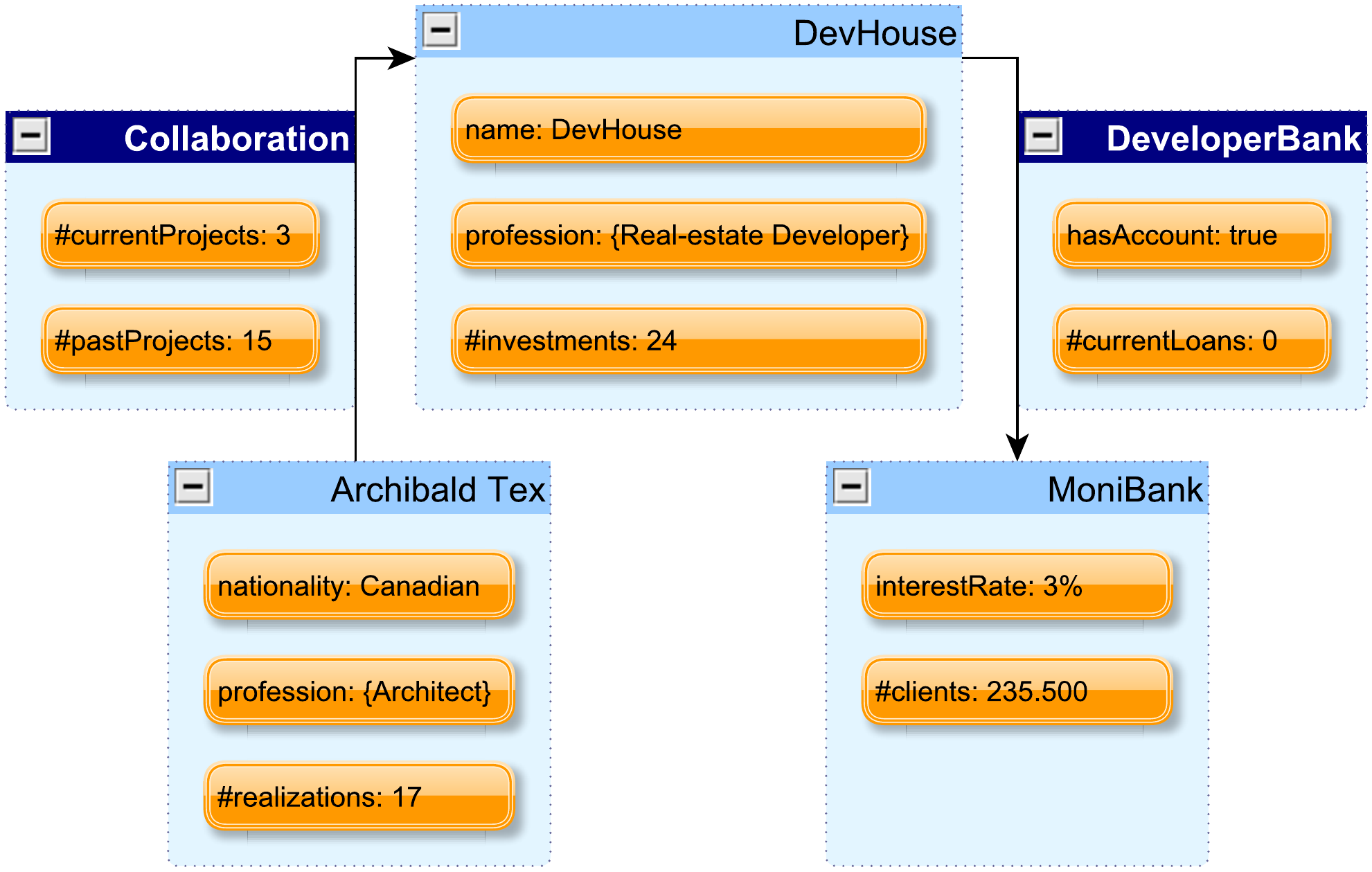}
	\caption{An example of an object-based graph}
	\label{fig:ObjectGraph}
\end{figure}

A simple object-based graph---illustrated in Figure~\ref{fig:ObjectGraph}---may consist of the \AT, \DHtt, and \MoniBank nodes. Nodes \AT and \DHtt are connected by an arc modelling the \texttt{Col\-lab\-o\-ra\-tion} between the architect and the real-estate developer defined as:
\begin{itemize}
	\item $\langle\; \texttt{\#currentProjects, 3}\; \rangle$, and
	\item $\langle\; \texttt{\#pastProjects, 15}\; \rangle$.
\end{itemize}
The second arc---\texttt{DeveloperBank}---connects \DHtt and \MoniBank.

In Figure~\ref{fig:ObjectGraph}, the three nodes \AT, \DHtt, and \MoniBank, are represented by rectangles, while the arcs between them are represented by arrows. Objects describing the arcs, \ie \texttt{Col\-lab\-o\-ra\-tion} and \texttt{DeveloperBank}, are represented by rectangles stuck to the arrows. The properties of both the nodes and the objects describing the arcs are represented by inner rounded rectangles.

%%%%%%%%%%%%%%%%%%%%%%%%%%%%%%%%%%%%%%%%%%%%%%%%%%%%%%%%%%%%%%%%%%%%%%%%%%%%%%%%%%%%%%%%%%%%%
%%%%%%%%%%%%%%%%%%%%%%%%%%%%%%%%%%%%%%%%%%%%%%%%%%%%%%%%%%%%%%%%%%%%%%%%%%%%%%%%%%%%%%%%%%%%%
\subsubsection{Class-based Graphs}

A \emph{class} is a set of property constraints $c=\{p^\alpha\}$. A \emph{property constraint} $p^\alpha$ is a pair $\langle n, \vartheta_n \rangle$, where $n$ is the name of the properties that $p^\alpha$ may constrain, and $\vartheta_n$ is a predicate.

As an example, consider a class \ExpArchClass, defined by the following set of property constraints:
\begin{itemize}
	\item $\langle\; \texttt{profession, $\supset$~\{Architect\}}\;\rangle$, and 
	\item $\langle\; \texttt{\#realizations, >15}\;\rangle$.
\end{itemize}

\begin{figure}[htp]
	\centering
		\includegraphics[scale=0.6]{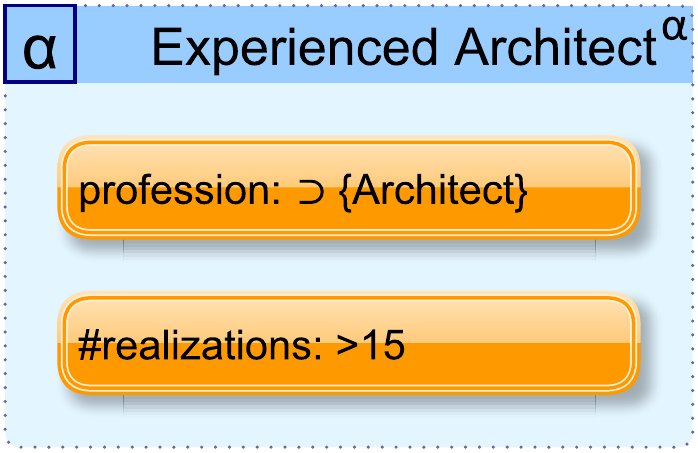}
	\caption{An example of a class}
	\label{fig:ClassExpArch}
\end{figure}

The class \ExpArchClass is illustrated in Figure~\ref{fig:ClassExpArch}. The external rectangle represents the class \ExpArchClass, while each inner rounded rectangle represents a property constraint of this class. Note the use of the '$\alpha$' Greek letter to mark class-related entities.

A set of classes and their relations may be modelled as a \emph{class-based graph}. 

\begin{definition}{Class-based Graph}
A class-based graph $g^\alpha= \langle N^\alpha,\vec{A}^\alpha \rangle$ is a graph whose nodes are classes, and arcs connecting nodes consist of a source class, a destination class, and a class describing the arc itself.
\end{definition}

To illustrate the concept of a class-based graph, consider a second class defining an \ExpDevClass by the following set of properties:
\begin{itemize}
	\item $\langle\; \texttt{profession, $\supset$~\{Real-estate Developer\}}\; \rangle$, and 
	\item $\langle\; \texttt{\#investments, >10}\; \rangle$.
\end{itemize}

\begin{figure}[htp]
	\centering
		\includegraphics[scale=0.6]{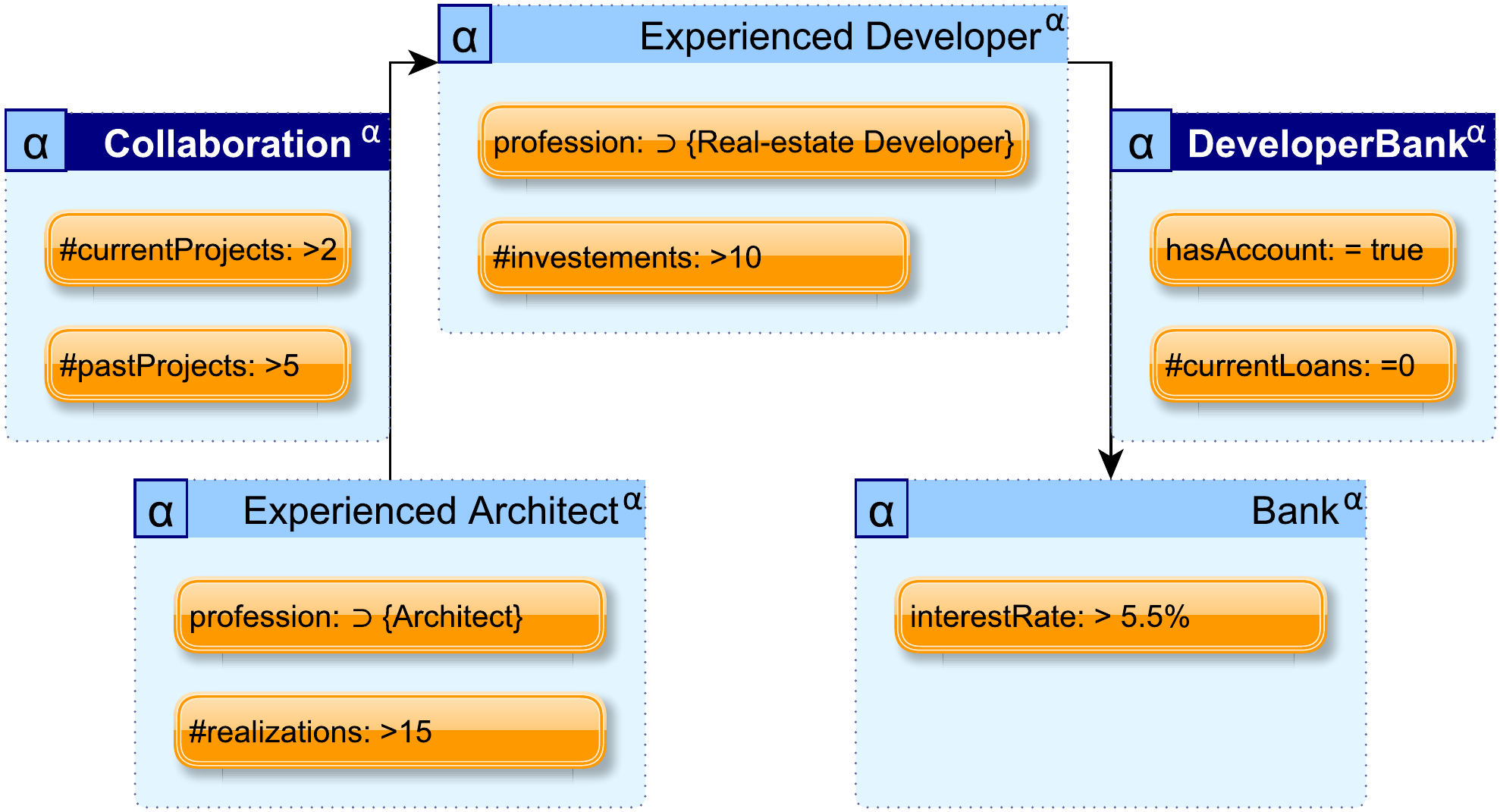}
	\caption{An example of a class-based graph}
	\label{fig:ClassGraph}
\end{figure}

A simple class-based graph---illustrated in Figure~\ref{fig:ClassGraph}---may consist of the \ExpArchClass, \ExpDevClass and \BankClass nodes. The nodes \ExpArchClass and \ExpDevClass are connected by an arc associated with the class \texttt{Collaboration$^\alpha$} defined by the following set of properties:
\begin{itemize}
	\item $\langle\; \texttt{\#currentProjects, >2}\; \rangle$, and
	\item $\langle\; \texttt{\#pastProjects, >5}\; \rangle$.
\end{itemize}
The second arc---\texttt{\DevBankClass}\unskip---connects \ExpDevClass and \BankClass.

In Figure~\ref{fig:ClassGraph}, the three nodes \ExpArchClass, \ExpDevClass and \BankClass, are represented by rectangles, while the arcs between them are represented by arrows. Classes describing the arcs, \ie \texttt{Col\-lab\-o\-ra\-tion$^\alpha$} and \texttt{\DevBankClass}, are represented by rectangles stuck to the arrows. The properties of both the nodes and the classes describing the arcs are represented by inner rounded rectangles. Note once again the use of the '$\alpha$' Greek letter to mark class-related entities.

An object $o=\bigl\{p=\langle n,v_n \rangle\bigr\}$ is an \emph{instance} of a class $c=\bigl\{p^\alpha=\langle n,\vartheta_n \rangle\bigr\}$, denoted $o \sqsubset c$,  iff all the property constraints of the class $c$ are satisfied by the properties of the object $o$. 
A property $p=\langle n,v_n \rangle$ \emph{satisfies} a property constraint $p^\alpha=\langle n', \vartheta_{n'} \rangle$, denoted $p \succ p^\alpha$, iff the name of the property and the property constraint are identical, \ie $n=n'$, and the predicate is true for the value of the property, \ie $\vartheta_{n'}(v_n)=\texttt{true}$.

Formally, $$o \sqsubset c \quad\Leftrightarrow\quad \forall p^\alpha \in c, \exists p \in o : p \succ p^\alpha.$$

The object \AT is an instance of the class \ExpArchClass because all the property constraints of the class are satisfied: \AT\ is an architect and his number of realizations, \ie 17, is higher than the required number, \ie 15. Note that additional properties, \eg the \texttt{nationality}, are not relevant for the class \ExpArchClass.

The object \AT is not an instance of the \ExpDevClass class for two reasons. First, the \texttt{profession} property constraint of \ExpDevClass is not satisfied by $\langle \texttt{profession, Ar\-chi\-tect} \rangle$. Second, no property named \texttt{\#investments} is even defined for \AT.

 \subsection{Processes}

%In this section, definitions presented in quotes have been originally proposed in the Technical Report Document Number WFMC-TC-1011 entitled ``Terminology and glossary'' by the Workflow Management Coalition~\cite{standard:WFMC:terminology}.

%\subsubsection{Process}
A \emph{process} is a set of activities which realize a business objective or a policy goal in a structured manner. A \emph{process instance} is a single enactment of a process.

%\subsubsection{Activity}
An \emph{activity} is a ``piece of work that forms one logical step within a process''~\cite{standard:WFMC:terminology}.
An activity may be automated work performed by information systems, \eg creating invoices by a Web service , or work performed by humans, \eg making a decision by a senior executive.

A \emph{state} is a ``representation of the internal conditions defining the status of a process instance at a particular point in time''~\cite{standard:WFMC:terminology}. In its simplest form, a state may be reduced to a label, \eg \texttt{item produced}.

%\subsubsection{Process Model}
\begin{definition}{Process Model}
A \emph{process model} $\pi^\alpha$ is a triplet $\langle S, A, \chi \rangle$, where $S$ is a set of states, $A$ is a set of activity descriptions, and $\chi$ is a relation $\chi : A \times S$.

The $\chi$ relation captures the possibility to execute a given activity in a given state: the activity described by $a \in A$ may be executed in the state $s \in S$ iff $a \chi s$.
\end{definition}

%Such a generic definition of process models do not limit social protocols to a given process modeling language or notation. Therefore, various process modeling languages or notations, such as BPMN~\cite{standard:OMG:BPMN12}, BPEL~\cite{standard:BPEL20}, Event-driven Process Chains (EPC)~\cite{epc:keller:1992}, or Petri nets~\cite{vda:1996:TGR}, may be used to model processes in social protocols. 

\subsection{Services}
%Except when explicitly stated differently, all service-related definitions have been originally proposed in the documents entitled ``Reference Model for Service Oriented Architecture 1.0''~\cite{standards:OASIS:SOA:referenceModel} and ``Reference Architecture Foundation for Service Oriented Architecture Version 1.0''~\cite{standards:OASIS:SOA:referenceArchitecture} proposed by the Organization for the Advancement of Structured Information Standards (OASIS).\vspace{1em}

An \emph{actor} is an entity (human or non-human) or organization of entities that is capable of action.

A \emph{need} is a measurable requirement that an actor is actively seeking to satisfy.

A \emph{capability} is an ability to achieve an effect.

\begin{definition}{Service}
A \emph{service }is an access to a capability of an actor, called a \emph{service provider}, to satisfy a need of a second actor, called a \emph{service consumer}, where the access is provided via a prescribed interface~\cite{standards:OASIS:SOA:referenceModel, standards:OASIS:SOA:referenceArchitecture}.
%A \emph{service} is an access to a capability owned by an actor ---~a \emph{service provider}~--- satisfying a need of other actors ---~\emph{service consumers}, where the access is provided using a prescribed interface.
%Services are the mechanism by which needs and capabilities are brought together. [\ldots] A \emph{service} is a mechanism to enable access to one or more capabilities, where the access is provided using a prescribed interface.
\end{definition}

A \emph{service interface} is the means for consuming a service.
It is admitted that an actor may provide a service to itself.

%\begin{definition}{Service Consumer}
%A \emph{service consumer} is an actor that interacts with a service in order to realize the real world effect produced by a capability to address a consumer need.
%\end{definition}
%
%\begin{definition}{Service Provider}
%A \emph{service provider} is an actor that offers a service that enables some capability to be used by other actors.
%\end{definition}

%% file: section_FormalModel.tex
\section{Formal Model of Service Protocols}
\label{sec:formalModel}

In this section, a formal model of service protocols is presented. First, the main elements of service protocols are introduced in an informal overview. Next, service-oriented summaries, service network schemata, and service protocols are formally defined.

\subsection{Overview}

Besides a process model defining the sequences of activities that may be performed during a collaboration process, a service protocol contains additionally two elements: a \emph{service-oriented summary} and a \emph{service network schema}.

A service-oriented summary provides a representation of the activities of an associated process model in SOA terms. The goal of the service-oriented summary of a service protocol is to represent activities of the process model as services, with each service consisting of a service consumer, a service interface, and a service provider. A service-oriented summary provides an abstraction of the activities as services independently of the language used to model the process, \eg BPEL or BPMN, focusing on the links between service consumers, providers, and interfaces.

A service network schema is a class-based graph that restricts the set of potential service elements, \ie service consumers, interfaces, and providers, that may participate in the service protocol by defining constraints on nodes and constraints on arcs, \ie social requirements. The goal of the service network schema of a service protocol is to define the set of collaborators that is required to execute the associated process instance.
\newline

As regards requirements for service protocols articulated in section~\ref{sec:requirements}, service-oriented summaries address partially the second requirement---separation of activities implementation from service protocols---by providing a means to describe activities in an abstract manner, independently of the underlying process model language. Service network schema addresses the fourth requirement---support for social aspects in collaboration---as a means to capture social requirements concerning the collaborators. The third requirement---strong mathematical foundations---is tackled by the formal model of service protocols presented below. 
\newline

The first ---reusability---and the second requirement---separation of activities implementation from service protocols---are addressed by providing four layers for service protocols: \emph{abstract}, \emph{prototype}, \emph{executable service protocols}, and \emph{service protocol instances}. An abstract service protocol does not provide any information concerning the implementation of the services. A prototype service protocol provides a partial implementation of the services defined in its service-oriented summary. An executable service protocol provides a complete implementation of the services defined in its service-oriented summary and may be instantiated.

The implementation of services encompasses implementations of service pro\-vi\-ders, service interfaces, and service consumers. In the proposed approach, it is assumed that a \emph{service network} is available as the source of service implementation used to build and instantiate executable service protocol. A service network is a network whose nodes are service providers, service interfaces, and service consumers.

\subsection{Service-Oriented Summaries of a Process Model}
%All abstract social protocols are based on a service-oriented summary (SoS) of the process model. In SoS, all activities that may potentially performed during the execution of the process are represented by triplets defining the ``who'' (the \emph{activator}), ``what'' (the \emph{description of the piece of work to be performed}), and ``whose'' (the \emph{activated}) part of the activity. 
%
%From a service-oriented perspective:
%\begin{itemize}
%	\item an activator is a \emph{class of service consumers}, 
%	\item a description of the piece of work to be performed is a \emph{class of service interfaces}, 
%	\item an activated is a \emph{class of service providers}. 
%\end{itemize}

In service protocols, the set of activities and states associated with a given type of human interactions are defined in a process model. No assumption is made on the chosen process modelling language or representation.

The idea underlying service-oriented summaries is to provide a common representation of activities in a process model, independently of the chosen process modelling language or representation, in terms of service consumers, service interfaces, and service providers.

In a service-oriented summary of a given process, all the activities that may potentially be performed during the execution of this process are represented by triplets defining the ``who'' (the \emph{service consumer}), ``what'' (the \emph{service interface}), and ``whose'' (the \emph{service provider}) part of the activity. These triplets are referred to as \emph{service descriptions}.

\begin{definition}{Service Description}
A \emph{service description} is a triplet $s_d^\alpha = \langle \; \textrm{sc}^{\,\alpha}, \;\; \textrm{si}^{\,\alpha}, \\ \textrm{sp}^{\,\alpha} \rangle \in S^\alpha_d$, where $\textrm{sc}^{\,\alpha}$ is a class of service consumers, $\textrm{si}^{\,\alpha}$ is a class of service interfaces, and $\textrm{sp}^{\,\alpha}$ is a class of service providers.
\end{definition}

A class of service consumers may for instance be the \ExpArchClass class defined in the illustrative example from Section~\ref{subsec:example}. Following on this example, a class of service interfaces may define access to printing services with constraints such as $\langle$\texttt{CAD plotting support}, $\supset \{$\texttt{bond, vellum}$\}\rangle$, $\langle$\texttt{pay\-ment means}, $\supset \{$\texttt{bank transfer, credit card}$\}\rangle$. Similarly, a class of service providers may define construction printing companies with constraints concerning the industry sector, the geographical location, etc.

Based on the definition of service description, a service-oriented summary of a process model may be defined as follows.

\begin{definition}{Service-Oriented Summary of a Process Model}
A \emph{service-orien\-ted summary} $\pi^\alpha_{\textsl{sos}}$  is a triplet $\langle \pi^\alpha, S^\alpha_d, \rho \rangle$, where $\pi^\alpha$ is a process model, $S^\alpha_d$ is a set of service descriptions, and $\rho: A \rightarrow S^\alpha_d$ is a function mapping activity descriptions in $\pi^\alpha$ to service descriptions in a bijective manner, \ie $\forall s^\alpha_d \in S^\alpha_d, \; \exists!a \in A, \textrm{such that } \rho(a)=s^\alpha_d$.
\end{definition}

Note that the only constraint on the process modelling language is the possibility to associate service descriptions with the activities to be performed in the associated process model. As a consequence, various process modelling languages, \eg BPEL, BPMN, may be used to model processes further summarized by service-oriented summaries.

\begin{figure}[htp]
	\centering
		\includegraphics[width=330pt]{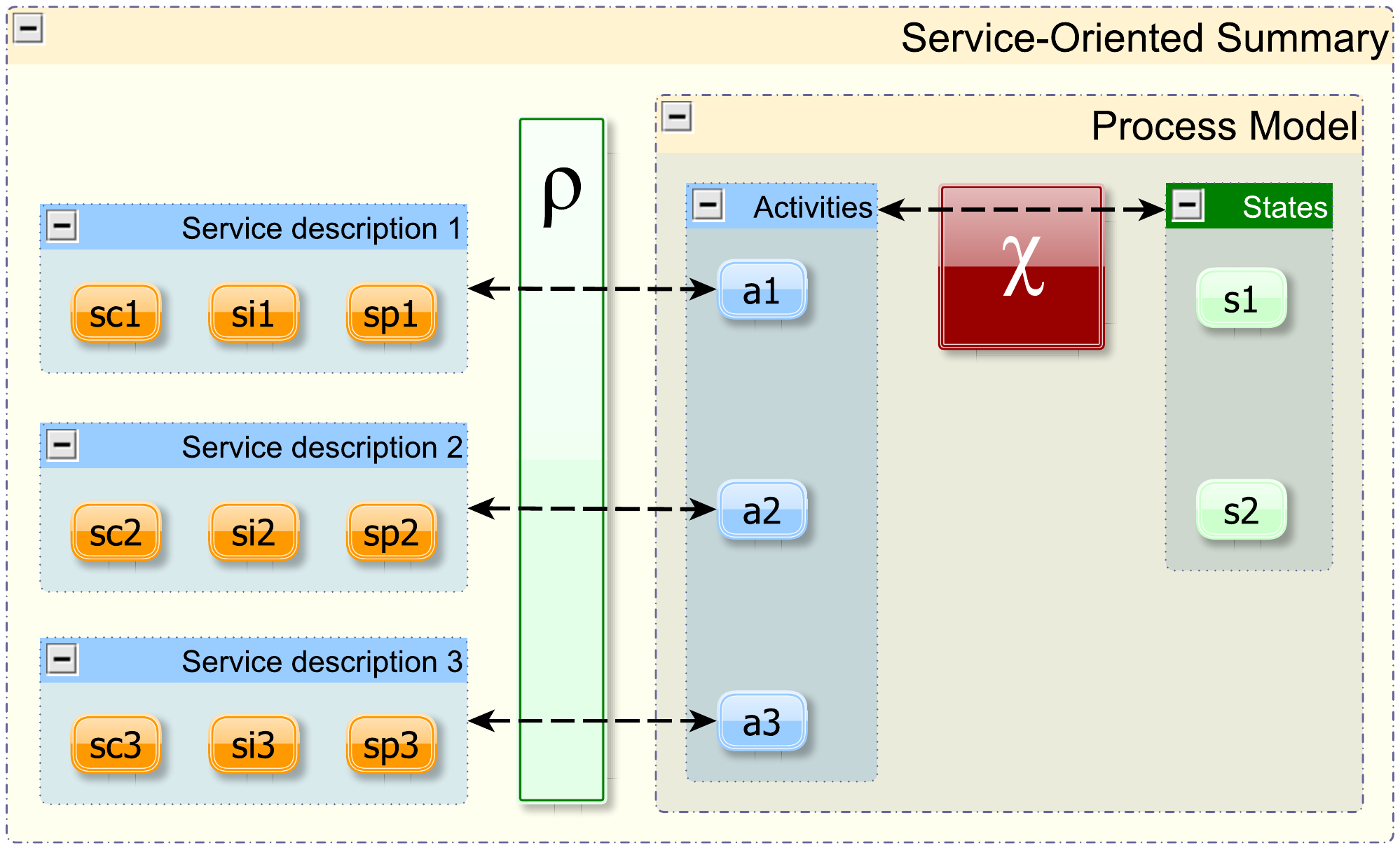}
	\caption{A service-oriented summary of a process model}
	\label{fig:SOS}
\end{figure}

The concept of service-oriented summary of a process model is illustrated in Figure~\ref{fig:SOS}, in which three service descriptions are represented as rectangles on the left side. Each service description contains a class of service consumers, represented by a rounded rectangle labelled \texttt{sc\{i\}}, a class of service interfaces, represented by a rounded rectangle labelled \texttt{si\{i\}}, and a class of service providers, represented by a rounded rectangle labelled \texttt{sp\{i\}}, where \(i \in {1,2,3}\). Next, three dashed arrows are connecting service descriptions with the activities of the process model represented by rounded rectangles labelled \texttt{a1}, \texttt{a2}, and \texttt{a3}. These three arrows visualize the mapping function \(\rho\). Besides the activities, the process model of the service-oriented summary contains a set of states, represented on the right side of Figure~\ref{fig:SOS}. The dashed arrows between the activities and the states represent the $\chi$ relation that captures the possibility to execute a given activity in a given state.

\subsection{Service  Network Schemata}

Service networks aim at capturing properties and relations among service entities, \ie service consumers, service interfaces, and service providers.
Service network schemata aim at capturing classes of service entities and their relations. Using an analogy with concepts from object-oriented programming, service network schemata may be considered as classes defining some ``templates'', while service networks may be considered as objects, each service networks consisting of its own ``state'' being its set of service entities.

\subsubsection{Service Networks and Schemata}

%\begin{definition}{Network}
%A \emph{network} is a linked set of entities.
%\end{definition}
%
%A network may be \emph{represented} by a graph, in which entities are represented by nodes, and the links between entities by arcs.
%
%\begin{definition}{Service  Entity}
%A \emph{service  entity} is an actor or a service interface.
%\end{definition}

\begin{definition}{Service  Network}
A \emph{service  network} is a network of service entities. A service entity is an actor or a service interface.
\end{definition}

A service  network  may be represented by an object-based graph, in which service  entities are represented by object-based nodes and links by object-based arcs.

%To improve readability, no distinction is further made between a social network and its representation as an object-based graph. The terms ``social entities'' and ``links between social entities'' are referring respectively to ``object-based nodes'' and ``object-based arcs'' of object-based graphs.

%To improve readability, the terms ``service  entities'' and ``links between service  entities'' are further used with the meaning of ``object-based nodes'' and ``object-based arcs'' of object-based graphs, replacing a service  network by its representation as an object-based graph (except when explicitely stated differently).

To emphasize service orientation, the graph-related terms ``object-based no\-des'' and ``object-based arcs'' are further replaced by their service network-related terms, \ie ``service entities'' and ``links between service entities'', respectively.

%\begin{definition}{Entity Property}
%An \emph{entity property} is a property of a node representing an entity of a service  network.
%\end{definition}
%
%\begin{definition}{Link Property}
%A \emph{link property} is a property of an arc representing a link of a service  network.
%\end{definition}

\begin{definition}{Service  Network Schema}
A \emph{service  network schema} is a network in which entities are classes of service  entities, and links are classes of links between service  entities.
\end{definition}

%Nodes of a social network schema --~classes of social entities~-- may be represented as sets of property constraints, that members of a given class --~social entities~-- have to satisfy.
%
%Arcs of a social network schema --~classes of links~-- may be represented by triplets $<e^\alpha_{src}, e^\alpha_{dst}, P^\alpha>$ consisting of a source node $e^\alpha_{src}$, a destination node $e^\alpha_{dst}$, and a set of property constraints $P^\alpha=\{<n, \vartheta_n>\}$, that member of a given class --~links~-- have to satisfy.

A service  network schema may be represented by a class-based graph, in which classes of service  entities are represented by class-based nodes and classes of links between service  entities by class-based arcs. 

%To simplify notations, arcs in class-based graphs are denoted $l^\alpha=<e^\alpha_{src}, ,e^\alpha_{dst},P^\alpha>$, where $e^\alpha_{src}$ is a node representing a class of service  entities that is the value of the property named \texttt{src}, $e^\alpha_{dst}$ is a node representing a class of service  entities that is the value of the property named \texttt{dst}, and $P^\alpha$ is a set of properties of the arc except the \texttt{src} and \texttt{dst} properties.

%To improve readability, no distinction is further made between a social network schema and its representation as a class-based graph. The terms ``classes of social entities'' and ``classes of links between social entities'' are referring respectively to ``classes'' and ``arcs'' of class-based graphs.

%To improve readability, the terms ``classes of service  entities'' and ``classes of links between service  entities'' are further used with the meaning of ``class-based nodes'' and ``class-based arcs'' of class-based graphs, replacing a service  network schema by its representation as a class-based graph (except when explicitely stated differently).

To emphasize service orientation, the graph-related terms ``class-based nodes'' and ``class-based arcs'' are further replaced by their service network-related terms, \ie ``classes of service entities'' and ``classes of links between service entities'', respectively.

\subsubsection{Memberships}

Compliance of a service network with a service network schema has a global character, although it is based on the local concept of \emph{membership}. Membership refers to a particular type of relations that may exists between objects and classes in a service network and a service network schema. 

Although various types of membership may be defined in service networks and service network schemata,  \emph{class relational membership} and \emph{link class full membership} have to be defined for a further definition of compliance.

\begin{definition}{Class Relational Membership}
A service  entity $e$ is a \emph{relational member} of a class of service  entities $e^\alpha$, denoted $e \topbullet\subset e^\alpha$, iff 
%\begin{inparaenum}[(1)] 
\renewcommand{\labelenumi}{(\arabic{enumi})}
\begin{enumerate}
	\item  $e$ is an instance of $e^\alpha$, 
	\item for each class of links starting from $e^\alpha$ and associated with a class $c$, at least one link starting from $e$ is associated with an instance of class $c$, and 
	\item for each class of links leading to $e^\alpha$ and associated with a class $c$, at least one link leading to $e$ is associated with an instance of class $c$.
\end{enumerate}
%\end{inparaenum}
 
\end{definition}

\noindent Formally, 
$$
	e \topbullet\subset e^\alpha \; \Longleftrightarrow \;
	\begin{cases}
	1) & e \sqsubset e^\alpha,\\
  2) & \forall l^\alpha=\langle e^\alpha,e^\alpha_{\textsl{dst}}, c \rangle,\; \exists\, l=\langle e,e_{\textsl{dst}},o\rangle \;:\; o \sqsubset c,\\
  3) & \forall l^\alpha=\langle e^\alpha_{\textsl{src}},e^\alpha, c\rangle,\; \exists\, l=\langle e_{\textsl{src}},e,o\rangle \;:\; o \sqsubset c.	
	\end{cases}
$$

%\begin{figure}[htp]
%	\centering
%		\includegraphics[width=\textwidth]{figs/ClassRelationalMembership}
%	\caption{Example of class relational membership. \JGpl\  is a relational member of the class \Prog}
%	\label{fig:ClassRelationalMembership}
%\end{figure}

In the example presented in Figure~\ref{fig:ObjectGraph} and \ref{fig:ClassGraph}, the service entity \DHtt is 
a relational member of the class \ExpDevClass.

First, \DHtt  is a member of the class \ExpDevClass. 

Second, there is only one class of links starting from the class of entities \ExpDevClass: \DevBankClass. All the constraints---\texttt{hasAccount} and \texttt{\#currentLoans}---defined in the class associated with the \DevBankClass class of links are satisfied by the object associated with the link \DevBank.

Third, there is only one class of links leading to the class of entities \ExpDevClass: \CollabClass. All the constraints---\texttt{\#currentPro\-jects} and \texttt{\#pastProjects}---defined in the class associated with the \CollabClass class of links are satisfied by the object associated with the link \Collab.

Note that the class relational membership of a given service entity $e$ to a class of service entities $e^\alpha$ is established on the basis of 
\begin{inparaenum}[(\itshape 1\upshape)]
	\item the properties and property constraints of $e$ and $e^\alpha$, and 
	\item the properties and property constraints of the links and classes of links starting and leading from/to $e$ and $e^\alpha$. 
\end{inparaenum}
Therefore the service entity \AT (respectively \MoniBank) not being a instance of the class of service entities \ExpArchClass (respectively \BankClass) is irrelevant for class relational membership of \DHtt. 

%%%%%%%%%%%%%%%%%%%%%%%%%%%%%
%\color{red}
%Class relational membership is illustrated in Figure~\ref{fig:ClassRelationalMembership}. In the presented example, the service entity \JGpl\  is a relational member of the class \Prog. 

%First, \JGpl\  is a member of the class \Prog. 
%
%Second, there exist two classes of links starting from the class of entities \Prog: \texttt{Workplace}$^\alpha$ and \texttt{Colla\-bo\-rators}$^\alpha$.  The link \texttt{Workplace} starting from \JGpl\ is a left member of the class of links \texttt{Workplace}$^\alpha$. The link \texttt{Collaborators} starting from the service entity \JGpl\  is a left member of the class of links \texttt{Collaborators}$^\alpha$. 
%
%Third, there is only one class of links leading to the class of entities \Prog: \texttt{Supervision}$^\alpha$. The link \texttt{Supervision} leading to \JGpl\ is a right member of the class of links \texttt{Supervision}$^\alpha$. 

%Note the class relational membership of a given service entity $e$ to a class of service entities $e^\alpha$ is established on the basis of 
%\begin{inparaenum}[(\itshape 1\upshape)]
%	\item the properties and property constraints of $e$ and $e^\alpha$, and 
%	\item the properties and property constraints of the links and classes of links starting and leading from/to $e$ and $e^\alpha$. 
%\end{inparaenum}
%Therefore the service entity \texttt{Oracle} being not a member of the class of service entities \texttt{University} is irrelevant for class relational membership. 

%%%%%%%%%%%%%%%%%%%%%%%%%%%%%
%\color{black}

\begin{definition}{Link Class Full Membership}
A link $l=\langle e_{\textsl{src}}, e_{\textsl{dst}}, o\rangle$ is a \emph{full member} of the class of links $l^\alpha=\langle e^\alpha_{\textsl{src}}, e^\alpha_{\textsl{dst}}, c\rangle$, 
denoted $l \backbullet\subset^{\bullet} l^\alpha$, iff the source and destination service entities are instances of their respective classes of service entities, and the object associated with the link is an instance of the class associated with the class of links, \ie 
	\begin{equation*}
	l \backbullet\subset^{\bullet} l^\alpha \quad\Longleftrightarrow\quad e_{\textsl{src}} \sqsubset e^\alpha_{\textsl{src}} \quad\wedge\quad e_{\textsl{dst}} \sqsubset e^\alpha_{\textsl{dst}} \quad \wedge\quad  o \sqsubset c
	\end{equation*}
\end{definition}

In the example presented in Figure~\ref{fig:ObjectGraph} and \ref{fig:ClassGraph}, the link \DevBank is 
a full member of the class of links \DevBankClass.
First, \DHtt is an instance of the \ExpDevClass class.
Second, \MoniBank is an instance of \BankClass class.
Third, all the constraints---\texttt{hasAc\-count} and \texttt{\#currentLoans}---defined in the class associated with the \DevBankClass class of links are satisfied by the object associated with the link \DevBank.

Only class relational membership and link class full membership are defined in this paper, as the definition of other types of membership is out of the scope of this paper.

\subsubsection{Compliance with Service Network Schemata}

Based on the membership relations defined above, the concept of compliance with a service network schema may be defined. A service network is compliant with a service network schema if the constraints on the service entities and the social requirements among them, defined in a service network schema, are satisfied by a given service network. As presented formally below, various levels of compliance may be distinguished.

\begin{definition}{Compliance Relation}
Consider a service network schema $\sigma^\alpha =  \langle E^\alpha, \\ L^\alpha \rangle$ and a service network $\sigma = \langle E, L \rangle$. 
A \emph{compliance relation} $\leftmodels$ on $\sigma \times \sigma^\alpha$  is a relation such that

\begin{equation}
\label{eq:complianceRelation1}
\forall (e,e^\alpha) \in E\times E^\alpha, \quad e \leftmodels e^\alpha \Rightarrow e \topbullet\subset e^\alpha,
\end{equation}

\begin{align}
\label{eq:complianceRelation2}
\forall (l^\alpha=<e^\alpha_{\textsl{src}}, e^\alpha_{\textsl{dst}}, c>)\in L^\alpha,\qquad\qquad\qquad\qquad\nonumber\\
 \forall (e_{\textsl{src}},e_{\textsl{dst}})\in E\times E : e_{\textsl{src}} \leftmodels e^\alpha_{\textsl{src}},  e_{\textsl{dst}} \leftmodels e_{\textsl{dst}}^\alpha, \qquad\qquad\qquad\qquad\nonumber\\
 \exists (l=<e_{\textsl{src}}, e_{\textsl{dst}}, o>) \in L \;:\; l\backbullet\subset^{\bullet} l^\alpha,
 \end{align}

\begin{equation}
\label{eq:complianceRelation3}
\forall e^\alpha \in E^\alpha, \exists e \in E \;:\; e \leftmodels e^\alpha. 
\end{equation}
\end{definition}

First, the compliance of a service entity $e$ with a class of service entities $e^\alpha$ implies that that the service entity $e$ is a relational member of the class of service entities $e^\alpha$ (cf.~Eq.~\ref{eq:complianceRelation1}). Second, for each class of links $l^\alpha$ between two classes of service entities $e^\alpha_{\textsl{src}}$ and $e^\alpha_{\textsl{dst}}$, for each service entities $e_{\textsl{src}}$ and $e_{\textsl{dst}}$ being members of $e^\alpha_{\textsl{src}}$ and $e^\alpha_{\textsl{dst}}$, respectively, there exists a link between $e_{\textsl{src}}$ and $e_{\textsl{dst}}$ that is a full member of $l$ (cf. Eq.~\ref{eq:complianceRelation2}). Third, for each class of service entities, at least one service entity is compliant with the class (cf. Eq.~\ref{eq:complianceRelation3}).

%\begin{figure}[htp]
%	\centering
%		\includegraphics[width=\textwidth]{figs/Compliance}
%	\caption{Example of a service network partially compliant with a service network schema}
%	\label{fig:Compliance}
%\end{figure}

\begin{definition}{Compliance with a Service  Net\-work Sche\-ma}\label{def:complianceRelation}
A service  network $\sigma = \langle E, L \rangle$ is \emph{compliant} with a service  network schema $\sigma^\alpha = \langle E^\alpha, L^\alpha \rangle$, denoted $\sigma \leftmodels\sigma^\alpha$, iff there exists a compliance relation $\leftmodels$ on $\sigma \times \sigma^\alpha$.
\end{definition}

\begin{definition}{Partial Compliance Relation}
Consider a service network schema $\sigma^\alpha=<E^\alpha, L^\alpha>$ and a service network $\sigma = < E, L>$. 
A \emph{partial compliance relation} $\dashv$ on $\sigma \times \sigma^\alpha$  is a relation that satisfies only the conditions of equations~\ref{eq:complianceRelation1} and \ref{eq:complianceRelation2}, the third condition being relaxed.

\end{definition}

\begin{definition}{Partial Compliance with a Service  Net\-work Sche\-ma}\label{def:partialComplianceRelation}
A service  network $\sigma = < E, L>$ is \emph{partially compliant} with a service  network schema $\sigma^\alpha=<E^\alpha, L^\alpha>$, denoted $\sigma \dashv \sigma^\alpha$, iff there exists a partial compliance relation $\dashv$ on $\sigma \times \sigma^\alpha$.
\end{definition}

%\begin{definition}{Full Compliance Relation}
%A \emph{full compliance relation} $\leftModels: E \times E^\alpha$ is a compliance relation such
%
%\begin{equation}
%\label{eq:fullComplianceRelation1}
%\forall e \in E , \exists e^\alpha \in E^\alpha \;:\; e \leftModels e^\alpha 
%\end{equation}
%\vspace{-24pt}
%\end{definition}
%
%\begin{definition}{Full Compliance with a Service  Net\-work Sche\-ma}\label{def:fullComplianceRelation}
%A service  net\-work $\sigma = < E, L>$ is \emph{fully compliant} with a service  network schema $\sigma^\alpha=<E^\alpha, L^\alpha>$, denoted $\sigma \leftModels \sigma^\alpha$, iff there~exists a full compliance relation $\leftModels: E \times E^\alpha$.
%%By extension, full compliance with a social network schema is denoted with the $\leftmodels$ symbol.
%
%\end{definition}

\begin{figure}[htp]
	\centering
		\includegraphics[width=330pt]{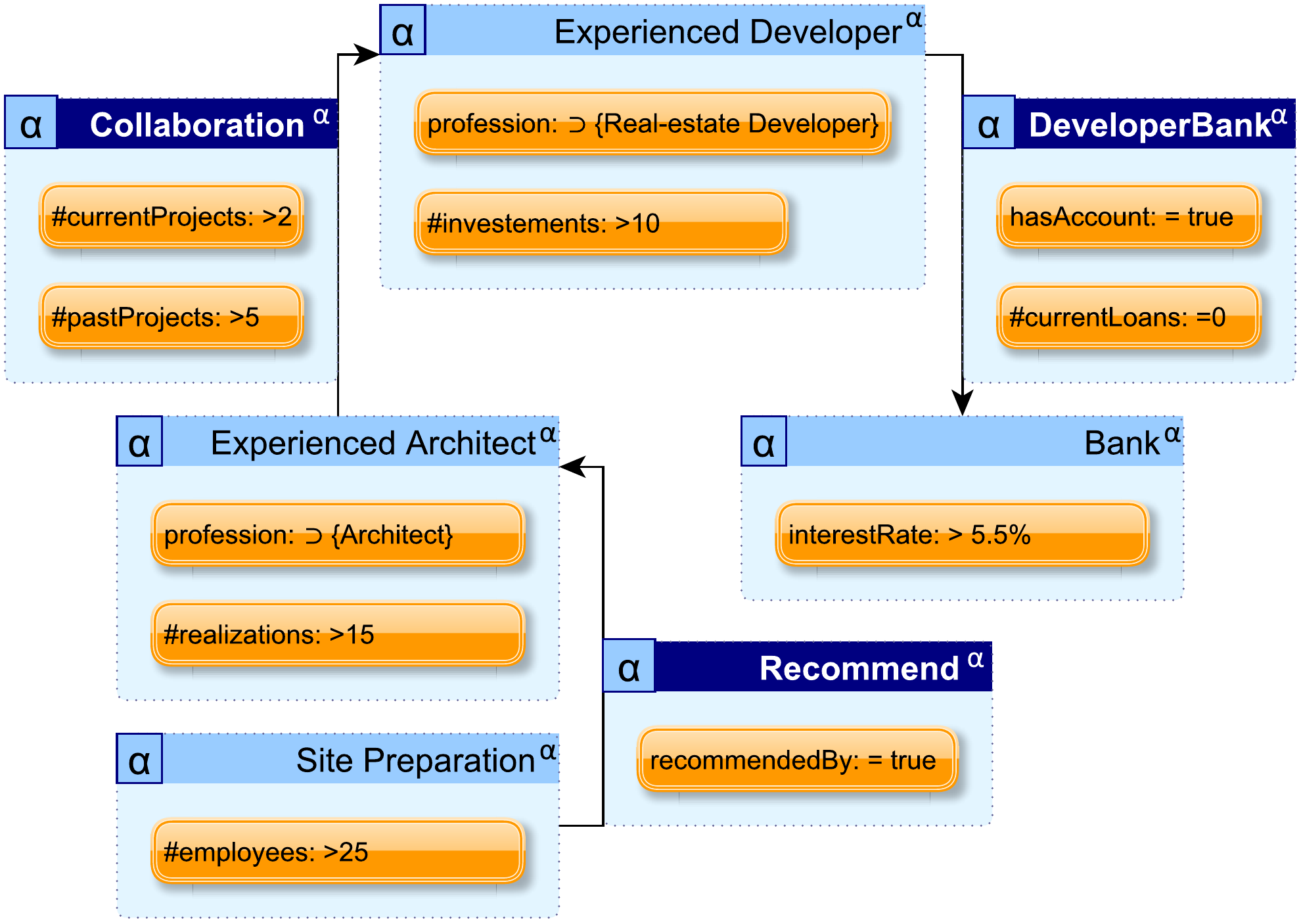}
	\caption{A service network schema with which the service network defined in Figure~\ref{fig:ObjectGraph} is partially compliant}
	\label{fig:ClassGraphExtended}
\end{figure}

Consider the service network schema presented in Figure~\ref{fig:ClassGraphExtended} to illustrate the concepts of partial compliance. This service network schema is an extension of the service network schema formerly presented in Figure~\ref{fig:ClassGraph}, with an additional class of service entities \SitePrepClass and an additional class of links \RecommendClass.

The compliance relation $\dashv$ applies to the following pairs of service entities and classes:
\begin{itemize}
	\item \AT $\dashv$ \ExpArchClass,
	\item \DHtt $\dashv$ \ExpDevClass, and
	\item \MoniBank $\dashv$ \BankClass.
\end{itemize}

Additionally, for each class of links between two classes of service entities among \ExpArchClass, \ExpDevClass, and \BankClass, there is a full member link between two service entities that are instances of the given classes. For example, the link \Collab is a full member of the class of links \CollabClass. As a consequence, both equations~\ref{eq:complianceRelation1} and \ref{eq:complianceRelation2} are satisfied by the relation  $\dashv$.

However, equation~\ref{eq:complianceRelation3} is not satisfied, as no service entity of the service network presented in Figure~\ref{fig:ObjectGraph} is an instance of the class of service entities \SitePrepClass.

As a conclusion, the relation $\dashv$ formerly defined is a partial compliance relation, and therefore, the considered service network is partially compliant with the service network schema presented in Figure~\ref{fig:ClassGraphExtended}.

\subsection{Service Protocols}

The concept of service protocol may be defined at four levels: 
\begin{itemize}
	\item at the \emph{abstract level}, a service-oriented summary provides a service-oriented representation of a process model, a service network schema provides constraints on service entities and social requirements, and additionally both the service-oriented summary and the service network schema are linked to associated service descriptions (from the service-oriented summary) with classes of service entities (from the service network schema);
	\item at the \emph{prototype level}, a service network is associated with both the service-oriented summary and the service network schema. At the prototype level, the service network provides only a partial implementation of an abstract service protocol, as some elements of the service-oriented summary and some classes of service entities of the schema may not be associated with any service entity of the service network;
	\item at the \emph{executable level}, the service network associated with both the service-oriented summary and the service network schema provides a complete implementation of an abstract service protocol: all the elements of the service-oriented summary and all the classes of service entities of the schema are associated with service entities of the service network;
	\item at the \emph{instance level}, an executable service protocol is enacted. At the instance level, service entities defined at the executable level are consuming and providing services modifying the state of the process model.
\end{itemize}

This four-level approach to human interactions answers directly the two first requirements presented in Section~\ref{sec:requirements}, \ie reusability and 
separation of activities implementation from service protocols.

\begin{definition}{Abstract Service Protocol}
An \emph{abstract service protocol} $\Upsilon^\alpha$ is a tri\-plet $\Upsilon^\alpha=<\pi^\alpha_{\textsl{sos}}, \sigma^\alpha, \aLambda >$, where $\pi^\alpha_{\textsl{sos}}$ is a service-oriented summary of a process model $\pi^\alpha$, $\sigma^\alpha=<E^\alpha, L^\alpha>$ is a graph representing a service  network schema, and $\aLambda$ is a mapping relation between the service-oriented summary and the service  network sche\-ma\footnote{The choice of the letter $\Upsilon$ for service protocols may be explained by the Greek word ``\greektext Uphres'ia\latintext'' (Ypires\'{i}a) which means ``service''.}.
%The choice of the letter $\Upsilon$ for social protocols may be explained by the greek word ``\greektext koinonik`oc\latintext'' (koinonikós) which means ``social''.

The mapping relation $\aLambda : (S\!\star^\alpha = SC^\alpha \cup SI^\alpha \cup SP^\alpha) \times E^\alpha$ associates elements of service descriptions---service consumer, service description and service provider classes---with classes of service  entities of the service  network schema.

$\forall (\textsl{s}\star^{\alpha}, e) \in S\!\star^\alpha \times E^\alpha, \;\;\textrm{s}\!\star^{\alpha} \aLambda\; e^\alpha$ means that the element of a service description $\textrm{s}\star^{\alpha}$---$\textrm{sc}^{\,\alpha}$, $\textrm{si}^{\,\alpha}$, or $\textrm{sp}^{\,\alpha}$---is associated with the class of service  entities $e^\alpha$ of the service  network schema.
\end{definition}

\begin{figure}[htp]
	\centering
		\includegraphics[width=\textwidth]{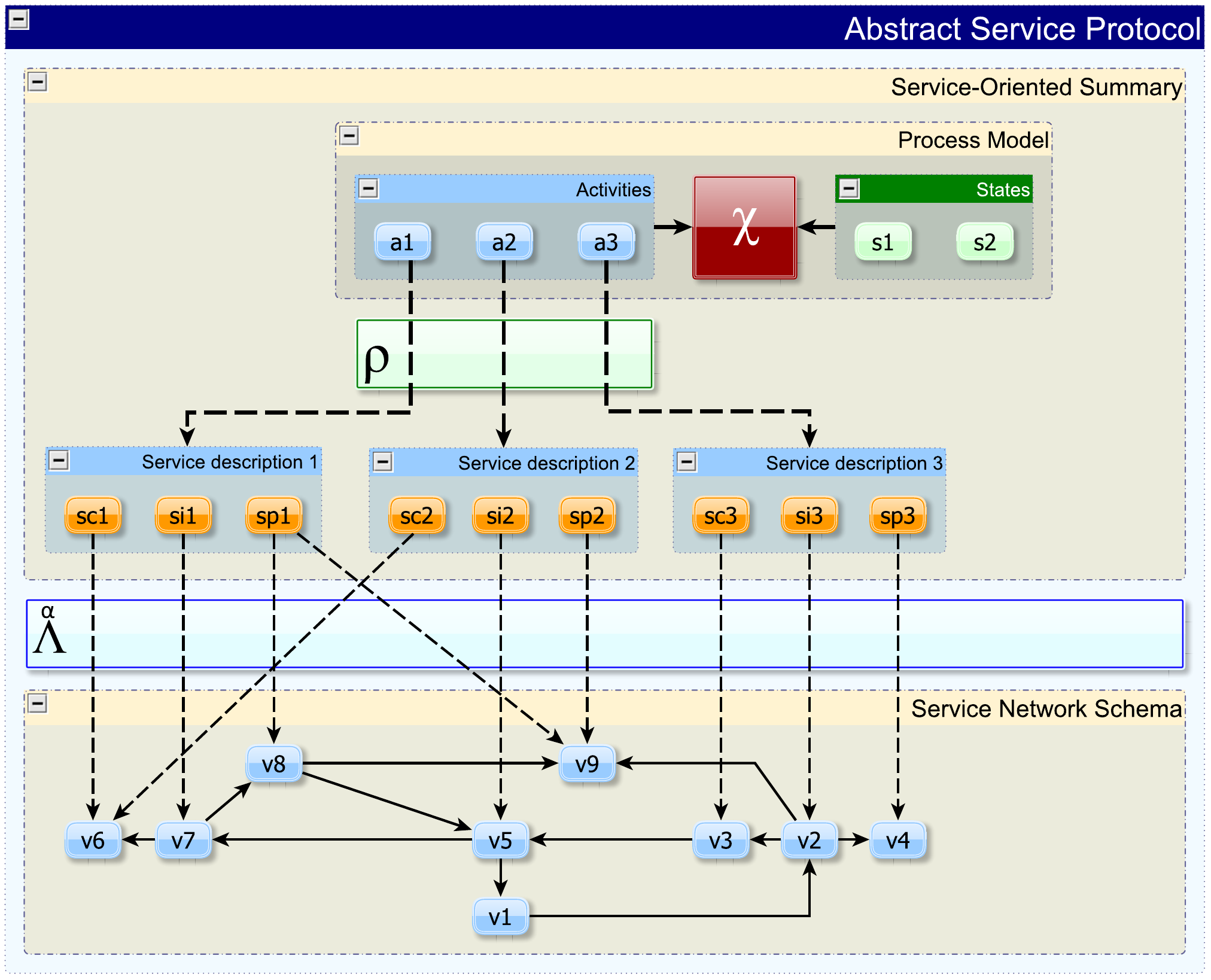}
	\caption{An abstract service protocol}
	\label{fig:AbstractSocialProtocol}
\end{figure}

An example of an abstract service protocol is illustrated in Figure~\ref{fig:AbstractSocialProtocol}. The service-oriented summary of the abstract service protocol is represented at the top of the Figure. The service network schema is represented at the bottom of the Figure. The representation of the service network schema has been simplified to a graph representation for the sake of readability. A set of dashed arrows associates the elements of service descriptions of the service-oriented summary with the nodes of the service network schema represented by rounded rectangles labelled \texttt{v\{i\}}, where \(i \in [1,9]\). This set of arrows represents the mapping relation \(\aLambda\) between the service-oriented summary and the service  network sche\-ma. Therefore, the service consumer \texttt{sc1} is associated with the class of service entities \texttt{v6} of the service network schema.

%\begin{figure}[htp]
%	\centering
%		\includegraphics[width=330pt]{figs/MappingSOS2SNS}
%	\caption{The mapping function $\Lambda$ linking the service-oriented summary (on the top) with the service  network schema (at the bottom)}
%	\label{fig:MappingSOS2SNS}
%\end{figure}

Additionally, the following relations among elements of service descriptions are defined:
\begin{enumerate}
	\item \textsl{sc} \texttt{<consumes, =~true>} \textsl{si},
	\item \textsl{sp} \texttt{<provides, =~true>} \textsl{si},
	\item \textsl{si} \texttt{<isConsumedBy, =~true>} \textsl{sc},
	\item \textsl{si} \texttt{<isProvidedBy, =~true>} \textsl{sp}.
\end{enumerate}

Although implicit in service descriptions, these relations are not explicitly defined in the service  network schema. However, the \texttt{provides} and \texttt{isProvidedBy} relations should be encompassed in service network schemata because these relations define social constraints on service interfaces and their provi\-ders. The extension of an existing service network schema to relations implicitly defined in service descriptions is an \emph{implicit service network schema}.

\begin{definition}{Implicit Service  Network Schema}
The implicit service  network  sche\-ma~$\sigma^{\alpha,+}=<E^{\alpha,+},L^{\alpha,+}>$ of an abstract service protocol is a supergraph of the service  network schema $\sigma^\alpha=<E^\alpha, L^\alpha,>$, \ie $\sigma^{\alpha,+} \supset \sigma^\alpha$, such that $E^{\alpha,+}=E^\alpha$ and $L^{\alpha,+}=L^\alpha \cup L^\alpha_\Lambda$, where $L^\alpha_\Lambda$ is defined as follows:
\begin{align*}
L^\alpha_\Lambda = &\\
%&\bigl\{<e_c^\alpha, e_i^\alpha, \texttt{<consumes, =~true>}> : (e_c^\alpha, e_i^\alpha) \in E^\alpha \times E^\alpha, \\
%&\parbox{5cm}{~}\exists s_d=<\textsl{sc}, \textsl{si}, \textsl{sp}> \in S_d : \textsl{sc} \;\Lambda\; e_c^\alpha \;\;\wedge\;\; \textsl{si} \;\Lambda\; e_i^\alpha\bigr\}
%\\
%&\bigcup
%\\
&\bigl\{<e_p^\alpha, e_i^\alpha, \texttt{<provides, =~true>}> : (e_p^\alpha, e_i^\alpha) \in E^\alpha \times E^\alpha, \\
&\parbox{5cm}{~}\exists s_d=<\textsl{sc}, \textsl{si}, \textsl{sp}> \in S_d : \textsl{sp} \;\aLambda\; e_p^\alpha \;\;\wedge\;\; \textsl{si} \;\aLambda\; e_i^\alpha\bigr\}
\\
&\bigcup
\\
%&\bigl\{<e_i^\alpha, e_c^\alpha, \texttt{<isConsumedBy, =~true>}> : (e_c^\alpha, e_i^\alpha) \in E^\alpha \times E^\alpha, \\
%&\parbox{5cm}{~}\exists s_d=<\textsl{sc}, \textsl{si}, \textsl{sp}> \in S_d : \textsl{sc} \;\Lambda\; e_c^\alpha \;\;\wedge\;\; \textsl{si} \;\Lambda\; e_i^\alpha\bigr\}
%\\
%&\bigcup
%\\
&\bigl\{<e_i^\alpha, e_p^\alpha, \texttt{<isProvidedBy, =~true>}> : (e_i^\alpha, e_p^\alpha) \in E^\alpha \times E^\alpha, \\
&\parbox{5cm}{~}\exists s_d=<\textsl{sc}, \textsl{si}, \textsl{sp}> \in S_d : \textsl{sp} \;\aLambda\; e_p^\alpha \;\;\wedge\;\; \textsl{si} \;\aLambda\; e_i^\alpha\bigr\}
\end{align*}
\end{definition}

\begin{figure}[htp]
	\centering
		\includegraphics[width=\textwidth]{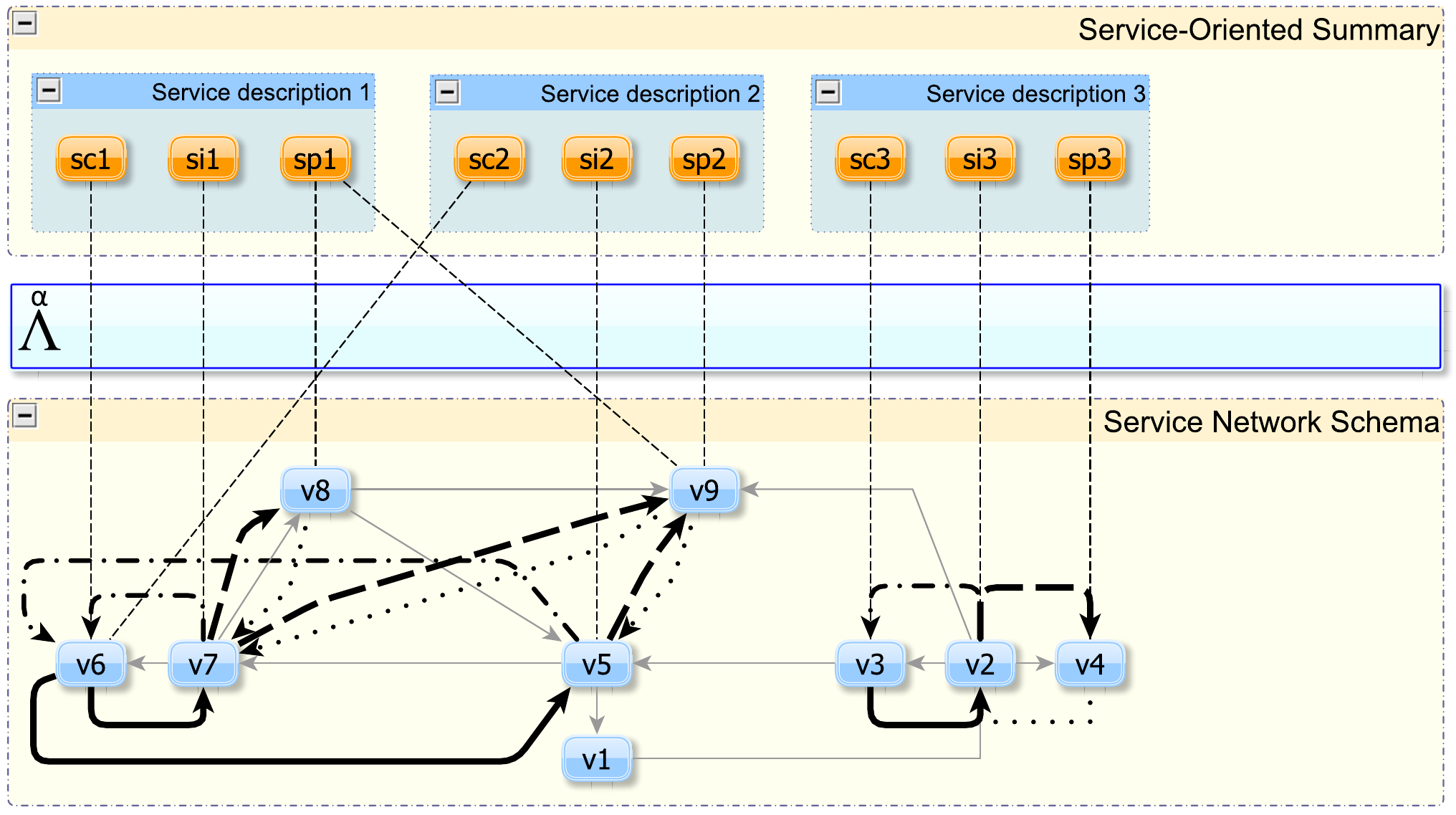}
	\caption{The implicit service  network schema for the service  network schema and the mapping function $\aLambda$ presented in Figure~\ref{fig:AbstractSocialProtocol}. Solid arrows between nodes represent \texttt{consumes} links. Dot-dashed arrows between nodes represent \texttt{isConsumedBy} links. Dashed arrows between nodes represent \texttt{isProvidedBy} links. Dot arrows between nodes represent \texttt{provides} links.}
	\label{fig:ImplicitSocialNetworkSchema}
\end{figure}

The implicit service network schema for the service network schema and the mapping function presented above is illustrated in Figure~\ref{fig:ImplicitSocialNetworkSchema}. On the top of the Figure, a simplified representation of a service-oriented summary is provided: only the service descriptions are represented. On the bottom of the figure, the service network schema already presented in Figure~\ref{fig:AbstractSocialProtocol} is represented, together with the links related with relations between the elements of service descriptions. Solid arrows between nodes represent \texttt{consumes} links. Dot-dashed arrows between nodes represent \texttt{isConsumedBy} links. Dashed arrows between nodes represent \texttt{isProvidedBy} links. Dot arrows between nodes represent \texttt{provides} links. The implicit service network schema is the service network schema formerly presented in Figure~\ref{fig:AbstractSocialProtocol} extended by all the links presented in Figure~\ref{fig:ImplicitSocialNetworkSchema}.

\begin{definition}{Prototype Service Protocol}
A \emph{prototype service protocol} $\Upsilon^\beta$ is a tuple $\Upsilon^\beta=<\Upsilon^\alpha, \sigma, \Omega, \Phi>$, where $\Upsilon^\alpha$ is an abstract service protocol, $\sigma$ is a service  network, and 
$\Omega$ is a relation $\Omega: E \times S\star^\alpha$ associating service  entities of the service  network to elements of service descriptions, and $\Phi: E \times E^\alpha$ associating nodes of the service  network to classes of service  entities of the service  network schema of $\Upsilon^\alpha$.

Let define the relation $\Phi^+: E \times E^{\alpha,+}$ as follows:
\begin{equation*}
\forall (e, e^{\alpha,+}) \in E\times E^{\alpha,+}, e\;\Phi^+\;e^{\alpha,+} 
\quad\Longleftrightarrow\quad 
	\begin{cases}
e\;\Phi\;e^{\alpha,+} \\
\vee \\
\exists s\!\star^\alpha \in S\star^\alpha :
(e \;\; \Omega \;\; s\!\star^\alpha) \wedge  (s\!\!\star^\alpha \aLambda\;\; e^{\alpha,+})
   \end{cases}
\end{equation*}

In prototype service protocols, the relation $\Phi^+$ ---~referred further as the \emph{induced service relation}~--- is a \emph{partial compliance} relation on $E \times E^{\alpha,+}$ (cf. Definition~\ref{def:partialComplianceRelation}).
\end{definition}

\begin{figure}[htp]
	\centering
		\includegraphics[width=330pt]{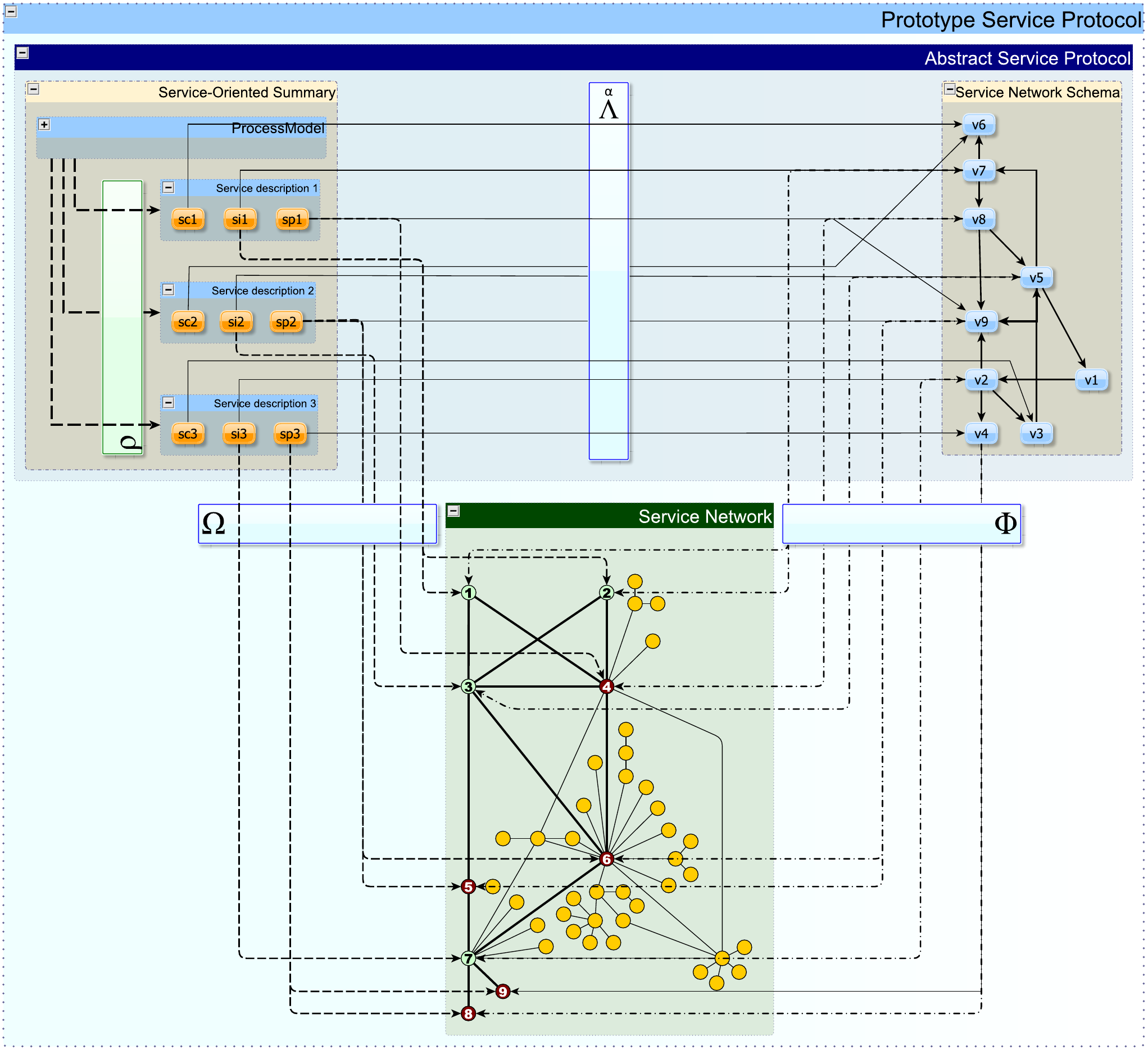}
	\caption{A prototype service protocol}
	\label{fig:PrototypeSocialProtocol}
\end{figure}

An example of a prototype service protocol is illustrated in Figure~\ref{fig:PrototypeSocialProtocol}. At the top, the abstract service protocol of the prototype service protocol is represented. At the bottom, the service network of the prototype service protocol is represented in a simplified manner as a graph. Each service entity of the service network is represented by a circle, some nodes being additionally labelled, \eg the top-right service network node labelled '1'. The links between service entities are represented by solid lines between the nodes.

The relation $\Omega$ mapping service entities of the service  network to elements of service descriptions is represented on the left side of the Figure by dashed arrows. As an example, the top-right service network node labelled '1' is associated with the service interface \texttt{si1}.

Note that many service entities may be associated with a given service description element. In Figure~\ref{fig:PrototypeSocialProtocol}, both nodes labelled '1' and '2' are associated with the service interface \texttt{si1}.

Additionally, the relation $\Phi$ mapping nodes of the service network to classes of service entities of the service  network schema is represented on the right side of the the Figure by dot-dashed arrows. As an example, the top-right service network node labelled '1' is associated with the class of service entities \texttt{v7}.

Note that many service entities may be associated with a given class of service entities of the service network schema. In Figure~\ref{fig:PrototypeSocialProtocol}, both nodes labelled '1' and '2' are associated with the class of service entities \texttt{v7}.

When a prototype service protocol is fully implemented, \ie there is a service entity implementing each element of service description, then it may be executed. The service protocol is then an executable service protocol.

\begin{definition}{Executable Service Protocol}
An \emph{executable service protocol} $\Upsilon^\epsilon$ is a prototype service protocol such that $\forall s\!\star^\alpha \in S\star^\alpha, \exists e \in E$ such that $e \;\Omega\; s\!\star^\alpha$, and the induced relation $\Phi^+:E \times E^{\alpha,+}$ is a \emph{compliance} relation (cf. Definition~\ref{def:complianceRelation}).
% (cf. Definition~\ref{def:complianceRelation}).
\end{definition}

\begin{definition}{Service Protocol Instance}
An \emph{instance of a service protocol} $\Upsilon$ is a pair $\Upsilon=(\Upsilon^\epsilon, s)$, where $\Upsilon^\epsilon$ is an executable service protocol, and $s$ is the current state of the associated process model.
\end{definition}

%% file: section_Applications.tex
\section{Potential Applications}
\label{sec:applications}

In this section, potential applications of service networks are presented in three areas: in the construction sector, in public administration, and in healthcare.

\subsection{Potential Applications in the Construction Sector}

In the construction sector, many investment processes concern complex construction works, such as the building of a mall, a warehouse, a highway, or a residential area. The realization of investment processes requires a multitude of varied competences usually provided by various organizations, such as architects, civil engineers, plumbing specialists, plant operatives, roofers, glaziers, and painters.

The heterogeneity of the set of activities involved in the execution of investment processes is accompanied by potentially complex precedence rules concerning the execution of the activities within a given investment process. As consequence, the execution of investment processes may be supported by appropriate process models specifying the precedence rules concerning the execution of activities.

Additionally, social aspects are playing an important role in the human interactions in the construction sector. Among various important social aspects, one may distinguish the history of former collaboration among organizations as a social aspect that usually influences the choice of collaboration partners. Similarly, the recommendation of an organization by another organization is a frequent situation in the construction sector, influencing here again the choice of collaboration partners. Trust is another social aspect having an important influence on the collaboration in the construction sector. As a consequence, the execution of investment processes, including the choice of collaboration partners, may be supported by encompassing the social network of the organizations in the construction sector in a given geographical area.

The encompassing of both process models and social networks leads to service protocols. Therefore, the execution of investment processes may be supported by service protocols, encompassing both process models and the social network of the organizations in the construction sector.

The four levels of abstraction of service protocols may be applied in the construction sector as follows:
\begin{itemize}
	\item An \emph{abstract service protocol} represents a given type of building investment projects and the rules associated with this type or, at least, part of such projects. To some extent, an abstract service protocol may document best practices that may potentially be reused by various organizations. As an example, an abstract service protocol may define the activities needed to lay down concrete foundations of a supermarket as a process model, \eg concrete stab pouring is followed by concrete levelling and drying. Next, the abstract service protocol defines the types of organization needed to perform these activities and the mandatory relationships between these types of organizations as a service network schema, \eg the architect should trust the concrete contractors;
	\item A \emph{prototype service protocol} represents the implementation of a given type of building investment projects by a given organization. As an example, a real-estate developer may already have some preferences concerning the concrete contractors they are collaborating with, even if the choice of the architect is not fixed. Another real-estate developer may use a different prototype service protocol based on the same abstract service protocol concerning foundations laying: he/she may have a partnering architect but no partnering concrete contractors;
	\item An \emph{executable service protocol} represents a complete implementation of a given type of building investment projects by a given organization. In an executable service protocol, all the organizations and services needed to implement the associated prototype service protocol are identified. As an example, consider an executable service protocol based on the prototype service protocol concerning the laying of foundations by the real-estate developer having a partnering architect. At the executable level, the concrete contractors are identified, as well as the services that they may  provide or consume during the execution of the associated abstract service protocol;
	\item A \emph{service protocol instance} supports the execution of a given building investment project or at least a part of it. As an example, the execution of the formerly presented executable service protocol leads to a service protocol instance executed by the partnering architect and the chosen concrete contractors. Besides the set of organizations and services, a service protocol instance has a state associated with the process model of the related abstract service protocol. As an example, a service protocol instance concerning foundations laying may be in the state ``concrete leveling''.
\end{itemize}

\subsection{Potential Applications in Public Administration}

In public administration, many administrative procedures concern complex administrative cases, such as the delivery of a construction permit, the processing of income taxes, the renewal of a passport. Similarly to the construction sector, the execution of administrative procedures requires various competences usually provided by various agencies, such as Internal Revenue Service (IRS), city's Department of Buildings, Passport Agencies.

Regulations constrain the competences of both clerks and agencies in administrative procedures. Regulations also often define precedence rules concerning the execution of the activities. As a consequence, the execution of administrative procedures may be supported by appropriate process models. In many public agencies, administrative procedures are already supported by IT systems.

Additionally, some administrative procedures define constraints concerning social aspects. A common type of social requirements found in administrative procedure concerns sibling relations. As an example, passport delivery for a child usually requires the presence of the child's parents or guardians at the agency at submission time. Many administrative procedures also require that the applicant lives in a particular geographical area related with the appropriate agency. As an example, the delivery of a construction permit is usually performed by a local agency whose jurisdiction with regard to construction permit delivery is geographically limited. As a consequence, the execution of administrative procedures may be supported by encompassing the social network of agencies and applicants, either natural or legal persons.

The encompassing of both process models and social networks in public administration leads to service protocols, in a similar manner as in the construction sector. Therefore, the execution of administrative procedure may be supported by service protocols, encompassing both process models and the social network of agencies and applicants.

The four levels of abstraction of service protocols in public administration may be applied as follows:
\begin{itemize}
	\item An \emph{abstract service protocol} represents a given administrative procedure or, at least, part of such a procedure. To some extent, abstract service protocol may serve as models of regulation concerning a given procedure. As an example, an abstract service protocol may define the activities needed to deliver a United States of America passport to a child as a process model, \eg the data---fulfilled forms, pictures, fingerprints---are gathered at a passport agency, next the data are checked, and then the passport is printed by the United States Government Printing Office. Next, the abstract service protocol defines the types of agencies and applicants needed to perform these activities and the mandatory relationships between them as a service network schema, \eg the child's parents or guardians have to fulfil and submit the forms;
	\item A \emph{prototype service protocol} represents an implementation of an  administrative procedure by a given agency. As an example, a given passport agency may have already some preferences concerning chosen aspects of the procedure, \eg the Buffalo Passport Agency does not require travel plans to be presented to apply for a passport. Another Passport Agency, such as the Western Passport Center, requires a proof of international travel within 2 weeks of the appointment or a proof of emergency abroad;
	\item An \emph{executable service protocol} represents a complete implementation of a given administrative procedure. Therefore, in an executable service protocol, all the agencies, applicants, and services needed to implement the associated prototype service protocol are identified. As an example, consider an executable service protocol based on the prototype service protocol concerning the delivery of a passport at the Buffalo Passport Agency. At the executable level, the clerk, the child, the parents or guardians are identified, as well as the services that they may  provide or consume during the execution of the associated abstract service protocol, such as providing the photography of the child or checking that the forms have been correctly fulfilled;
	\item A \emph{service protocol instance} supports the execution of an administrative procedure or, at least, a part of it. As an example, the execution of the formerly presented executable service protocol leads to a service protocol instance executed by the Buffalo Passport Agency. Besides the clerk, the child, the parents or guardians, and services, a service protocol instance has a state associated with the process model of the related abstract service protocol. As an example, a service protocol instance concerning passport delivery may be in the state ``passport waiting for printing by the United States Government Printing Office''.
\end{itemize}

\subsection{Potential Applications in Healthcare}

In healthcare, two types of processes may be distinguished. First, many procedures in healthcare are administrative procedures, \eg patient registration and scheduling, drugs management, medical record management. Second, key procedures in healthcare are medical procedures related with patient healing, \eg surgery operation, vaccination, broken bone immobilization.

These two types of processes share a set of common characteristics. First, the execution of activities of these processes often requires certified competences: only a pharmacist is allowed to manage drugs at a hospital, only anaesthesiologists---eventually nurse anaesthetists in some countries---are allowed to perform anaesthesia. Second, precedence rules are important for both administrative and medical procedure: a patient has to be registered before she/he may be examined by a physician, an X-ray radiography precedes the immobilization with a plaster or fibreglass cast. Third, the number of competences involved in the administrative and medical procedures is usually high.

Similarly to the processes in the construction sector and public administration, the execution of both administrative and medical procedures may be supported by appropriate process models. Support for administrative procedure by IT systems is already implemented in many healthcare facilities. Computer support for medical procedures is rapidly becoming popular with advanced imaging techniques and robotics. However, support for the process facet of medical procedures is still to be widely  implemented and accepted.

Additionally, social aspects are playing an important role in some healthcare procedures. First, administrative procedures may restrict relations between the patients and the healthcare facilities. As an example, in the Polish healthcare system, a patient has to consult a primary care physician, also known as family physician, that may further redirect the patient to a medical specialist. The primary care physician and the patient have to be located in the same geographical area. Second, a medical procedure may also constrain relations between persons involved in the procedure. As an example, consider a live donor transplantation. In this transplantation operation, part of the liver of an parent, sometimes a sibling, is transplanted to another person, usually a child. The parent and the child have to share the same blood type, the donor should be have a similar or bigger size than the recipient. As a consequence, the execution of healthcare procedures may be supported by encompassing the social network of patients and healthcare personnel.

The encompassing of both process models and social networks leads to service protocols. Therefore, the execution of healthcare procedures may be supported by service protocols, encompassing both process models and the social network of patients and the healthcare personnel.

The four levels of abstraction of service protocols may be applied as follows in healthcare:
\begin{itemize}
	\item An \emph{abstract service protocol} represents a given administrative or medical procedure. To some extent, abstract service protocol may serve to document best practices that may potentially be reuse by various healthcare facilities. As an example, an abstract service protocol may define the activities needed to perform a live donor liver transplantation, as well as the relations between the donor and the recipient;
	\item A \emph{prototype service protocol} represents the implementation of a healthcare procedure at a given healthcare facility. As an example, a hospital may already have some preferences concerning the accommodations of the donor and the recipient in adjacent rooms. Another hospital may have a different approach, separating donors from recipients;
	\item An \emph{executable service protocol} represents a comprehensive implementation of a healthcare procedure by a given facility. In an executable service protocol, all the patients, the medical personnel, and services needed to implement the associated prototype service protocol are identified. As an example, consider an executable service protocol based on the prototype service protocol concerning the live donor liver transplantation at the healthcare facility accommodating the donors and patients in adjacent rooms. At the executable level, the patients, the surgeon, the anaesthesiologist, and the nurses are identified, as well as the services that they may provide or consume during the execution of the associated abstract service protocol;
	\item A \emph{service protocol instance} represents the execution of a given healthcare procedure. As an example, the execution of the formerly presented executable service protocol leads to a service protocol instance executed by the patients and the healthcare personnel. A service protocol instance has a state associated with the process model of the related abstract service protocol. As an example, a service protocol instance concerning live donor liver transplantation may be in the state ``donor and recipient in post anaesthesia care unit''.
\end{itemize}

%% file: section_Discussion.tex
\section{Discussion}
\label{sec:discussion}

In this section, first, positive aspects of service protocols are presented. In a second part, limitations of service protocols are detailed.

Among positive aspects of service protocols, the support for social requirements has to be highlighted. First, the concept of social requirement is clearly defined in this paper as class of links between service entities. To our best knowledge, this is the first definition of social requirement in the SOA context. Second, service network schemata, in which social requirements and requirements concerning service entities themselves, are formally defined, based on graphs. The formalized model of service protocols is therefore based on fully formalized concepts relying on sound foundations (object orientation, graph theory, BPM, SOA). Additionally, sound foundations for service protocols would ease the  development of tools to design, manage, and validate service protocols.

Another interesting characteristic of service protocols is their service orientation.  With the wide adaptation of SOA, even if in many cases just at the Web service level, the underlying concepts of service protocols, \ie service consumer, service interface, service provider, are broadly accepted and understood. Therefore, the learning curve for service protocols should be smooth and not steep.

In the presented approach, the instantiation of service protocols assumes the existence of a service network in which a set of available service entities and their relations is represented. Many social websites are currently publishing personal data of private users and their relations.
Social networks has came to the public awareness with the recent success of social websites, such as Facebook~\cite{www:facebook}, Qzone~\cite{www:qzone}, Twitter~\cite{www:twitter}, MySpace~\cite{www:myspace}, LinkedIn~\cite{www:linkedin}, or Orkut~\cite{www:orkut}, just to name a few. With more than 800 million active users announced by Mark Zuckerberg during the F8 conference~\cite{www:facebook:f8} on September 22, 2011, the possibility to build large networks is demonstrated, and it is probably correct to assess that social or service networks will be a stable element of the future IT landscape. 

In the context of service networks, choosing service entities in a large service network is an interesting characteristic from the perspective of the \emph{adaptation of service protocols}. As in~\cite{picard:prove:2009}, the adaptation of service protocols refers in this paper to the capability of a group of human interacting according to a given service protocol to modify the service protocol at run-time to react to changing conditions, external or internal to the group of interacting humans. When a group has to adapt the service protocol ruling its interactions, a frequent task is the deletion/replacement/ad\-di\-tion of new collaborators/tools. Having in mind that more than half of all the employers (53\% according Holzer~\cite{holzer:1987}, 60\% according to Bewley~\cite{bewley:1999}) are seaking future employees on the social networks of their employees, adaptation of service protocols may take a serious advantage of the service network underlying service protocols.

Finally, a solution to the problem of the instantiation process for service protocols have already been proposed: a method for partner and service selection based on social protocols (which are a particular type of service protocols) is presented in~\cite{picard:prove:2010}. In this method, first, a set of service entities covering the functionalities required for the process to run is selected by selecting members of classes of service entities from a service network. Second, a set of potential groups of partners are created using a generic algorithm within with social requirements, expressed as classes of links among service entities, are used to evaluate and filter out inappropriate groups. The method presented in~\cite{picard:prove:2010} is fully implemented.

~

Among drawbacks, the lack of semantics associated with property names and property constraint names may be distinguished. It is highly probable that various service protocol designers are using different words to refer to the same property. As an example, a property name \texttt{myLanguages} may have some semantic relation, \eg be a synonym, with a property name \texttt{supportedLanguages}. A more complex case may be the semantic relation between \texttt{geographicalLocation} and \texttt{country}. Semantic relations may exist not only among property names, or among property constraint names, but among exist between a property name and a property constraint. Consider a property constraint $p^\alpha$ named \texttt{geographicalLo\-ca\-tion} with the value \texttt{ = Indonesia} and a property $p$ named \texttt{country} with the value \texttt{In\-do\-ne\-sia}. Without the semantic relation between \texttt{geographicalLoca\-tion} and \texttt{country}, $p$ does not satisfy $p^\alpha$. A support for semantics could improve the representation of classes of service entities, class of links between service entities, service entities, and links between service entities.

Another limitation of service protocols is its limited expressive power. In its current form, the expressivity of service protocols does not encompass constraints concerning neither the cardinality of class members, nor alternatives. Constraints concerning the cardinality would capture for example that in a given process more than 2, but less than 20, programmers are required, or that exactly 3 architects are required. Alternatives on classes may for example capture the fact that either a certified architect or a senior developer are needed in a given social protocol. Currently, service protocols do not support the specification of such requirements. 

Finally, an important limitation concerning service protocols is the lack of validated methodology concerning the design of service protocols. Various approaches to service protocol design may be considered: one may start by designing a process model, then a service-oriented summary may be built for the process model, next a service network schema may be designed, and finally, the service network schema may be linked with the service-oriented summary. The second approach may start with the specification of a service network schema, next service descriptions may be defined and associated with classes of service entities, and finally service descriptions may be associated with activities structured by a process model. To our best knowledge, methodologies for the design of service protocols are still to be developed.

%% file: section_Conclusions.tex
\section{Conclusions}
\label{sec:conclusions}

In this paper, the concept of service protocol has been presented and fully formalized. Service protocols are based on sound foundations in the areas of graph theory, SOA, and BPM. However, the combination of concepts from these areas is innovative as the proposed model goes far beyond recently proposed ideas such as Social BPM or ACM.

A major contribution to the modelling of human interactions is the introduction of service network schemata. The integration of service network schemata with process models leads to a model of human interactions supporting not only the definition of the potential sequences of interactions, but also the definition of the expected properties of actors and their relations required for the process to be executed. As a consequence, a service protocol may be considered as a model defining the ``composition'' of a group within which interactions are structured by a given process model. Service protocols capture social requirements for process models in service network schemata and the link between schemata and service-oriented summaries.

Another major contribution to the modelling of human interactions is a unified approach organized around the concept of service. It should be clearly stated that, in this paper, the word service has a broader meaning than just Web service, encompassing not only service provided by software entities, but also services provided by humans and organizations. As a consequence, in the proposed model, process models are summarized with service descriptions, social requirements are modelled as classes of service entities and classes of links among service entities. Note that the concept of service description which is directly associated with process activities is a triplet consisting of service consumers, service interfaces, and service providers. Such a representation of process activities is more general than representation of process activities in BPEL: in BPEL, a service may be reduced to its interface and its service provider, the service consumer being the BPEL engine. Process models in which the service consumer is actually consuming services may be supported by the proposed model. 

~

Among future works, performance and scalability of the proposed model is an issue that requires more efforts. Although checking if a service network is compliant with a service network schema is a relatively easy task, identifying within a large service network a service sub-network that is compliant with a service network schema is a complex task. To our best knowledge, algorithms to perform this search task efficiently are still to be developed.
Another topic in which further work is needed are semantic issues related to property names and property constraint names. In its current version, the potential semantic relations among property names, among property constraint names, and among property names and property constraint names are not taken into account.
Finally, although a pilot application has been developed, validating the feasibility of a prototype supporting the presented model, validation of the pertinence of this model requires an application in real-world cases. The pilot application is currently tailored to the needs of the construction sector with a scenario involving a real-estate developer enterprise and its subcontractors. In this business environment, social requirements play an important role: examples of recurrent social elements are frequent recommendation of one SME by another or cooperation history on past construction places.